\documentclass{amsart}

\usepackage[sortcites=true, maxnames=99, firstinits=true, citestyle=alphabetic, style=numeric, backend=bibtex]{biblatex}
\usepackage{amsthm}
\usepackage{amsmath}
\usepackage{amssymb}
\usepackage{hyperref}
\usepackage{cleveref}
\usepackage[disable]{todonotes}
\usepackage{ifthen}
\usepackage[inline]{enumitem}

\usepackage{tikz}
\usepackage{tikz-qtree}

\newcommand{\ratunion}{\cup}
\newcommand{\C}{\mathcal{C}}
\newcommand{\D}{\mathcal{D}}
\newcommand{\Dclosure}[1]{#1\mathord{\downarrow}}
\newcommand{\emptyWord}{\varepsilon}
\newcommand{\Lang}[1]{\mathsf{L}(#1)}
\newcommand{\Trans}[1]{\mathsf{T}(#1)}

\newcommand{\PLang}[1]{\mathsf{PL}(#1)}
\newcommand{\IW}[3]{\mathsf{IW}_{#1}(#2,#3)}
\newcommand{\Parikh}[1]{\Psi(#1)}
\newcommand{\alphabet}{\mathsf{alph}}

\newcommand{\N}{\mathbb{N}}

\newcommand{\subword}{\preceq}
\newcommand{\yield}[1]{\mathsf{yield}(#1)}

\newcommand{\ParikhMap}{\Psi}

\newcommand{\reverse}[1]{{#1}^{\mathsf{rev}}}
\newcommand{\Powerset}[1]{2^{#1}}

\newcommand{\autpath}[3]{#1\xrightarrow{#2}#3}
\newcommand{\transpath}[4]{#1 \mathrel{\raisebox{-2.5pt}{$\xrightarrow{#2,#3}$}} #4}
\newcommand{\transedge}[4]{(#1,#2,#3,#4)}
\newcommand{\autedge}[3]{(#1,#2,#3)}
\newcommand{\autstep}[1][YYY]{\ifthenelse{\equal{#1}{YYY}}{\rightarrow}{\rightarrow_{#1}}}
\newcommand{\autsteps}[1][YYY]{\ifthenelse{\equal{#1}{YYY}}{\rightarrow^*}{\rightarrow^*_{#1}}}
\newcommand{\autstepsn}[2][YYY]{\ifthenelse{\equal{#1}{YYY}}{\rightarrow^{#2}}{\rightarrow^{#2}_{#1}}}
\newcommand{\grammarstep}[1][YYY]{\ifthenelse{\equal{#1}{YYY}}{\Rightarrow}{\Rightarrow_{#1}}}
\newcommand{\grammarsteps}[1][YYY]{\ifthenelse{\equal{#1}{YYY}}{\Rightarrow^*}{\Rightarrow^*_{#1}}}
\newcommand{\grammarstepsn}[2][YYY]{\ifthenelse{\equal{#1}{YYY}}{\Rightarrow^{#2}}{\Rightarrow^{#2}_{#1}}}
\newcommand{\grammarstepp}[1][YYY]{\ifthenelse{\equal{#1}{YYY}}{\Rightarrow'}{\Rightarrow'_{#1}}}
\newcommand{\grammarstepsp}[1][YYY]{\ifthenelse{\equal{#1}{YYY}}{\Rightarrow'^*}{\Rightarrow'^*_{#1}}}
\newcommand{\grammarsteppp}[1][YYY]{\ifthenelse{\equal{#1}{YYY}}{\Rightarrow''}{\Rightarrow''_{#1}}}
\newcommand{\grammarstepspp}[1][YYY]{\ifthenelse{\equal{#1}{YYY}}{\Rightarrow''^*}{\Rightarrow''^*_{#1}}}
\newcommand{\grammarsteplin}[1][YYY]{\ifthenelse{\equal{#1}{YYY}}{\Rightarrow_{\mathsf{lin}}}{\Rightarrow_{#1,\mathsf{lin}}}}
\newcommand{\grammarstepslin}[1][YYY]{\ifthenelse{\equal{#1}{YYY}}{\Rightarrow_{\mathsf{lin}^*}}{\Rightarrow^*_{#1,\mathsf{lin}}}}

\newcommand{\iffSpace}{\quad}
\newlist{conditions}{enumerate}{1}
\setlist[conditions,1]{label=(\roman*)}
\crefname{conditionsi}{condition}{conditions}
\Crefname{conditionsi}{Condition}{Conditions}

\newtheorem{theorem}{Theorem}

\newtheorem{lemma}[theorem]{Lemma}
\newtheorem{corollary}[theorem]{Corollary}
\newtheorem{proposition}[theorem]{Proposition}
\newtheorem{example}[theorem]{Example}

\begin{document}

\title{An approach to computing downward closures}
\author{Georg Zetzsche}
\address{
Technische Universit\"{a}t Kaiserslautern \\
Fachbereich Informatik \\
Concurrency Theory Group}
\email{zetzsche@cs.uni-kl.de}

\begin{abstract}
The downward closure of a word language is the set of all (not necessarily
contiguous) subwords of its members. It is well-known that the downward closure
of any language is regular. While the downward closure appears to be a powerful
abstraction, algorithms for computing a finite automaton for the downward
closure of a given language have been established only for few language
classes.

This work presents a simple general method for computing downward closures.  
For language classes that are closed under rational transductions, it is shown
that the computation of downward closures can be reduced to checking a certain
unboundedness property.

This result is used to prove that downward closures are computable for
\begin{enumerate*}[label=(\roman*)]\item every language class with effectively
semilinear Parikh images that are closed under rational transductions, \item
matrix languages, and \item indexed languages (equivalently, languages accepted
by higher-order pushdown automata of order~2).\end{enumerate*}
\end{abstract}

\maketitle

\section{Introduction}
The \emph{downward closure} $\Dclosure{L}$ of a word language $L$ is the set of
all (not necessarily contiguous) subwords of its members. While it is
well-known that the downward closure of any language is
regular~\cite{Haines1969}, it is not possible in general to compute them.
However, if they are computable, downward closures are a powerful abstraction.
Suppose $L$ describes the behavior of a system that is observed through a lossy
channel, meaning that on the way to the observer, arbitrary actions can get
lost. Then, $\Dclosure{L}$ is the set of words received by the
observer~\cite{HabermehlMeyerWimmel2010}.  Hence, given the downward closure as
a finite automaton, we can decide whether two systems are equivalent under such
observations, and even whether one system includes the behavior of another.

Further motivation for studying downward closures stems from a recent result of
\citeauthor{CzerwinskiMartens2014}~\cite{CzerwinskiMartens2014}. It implies that
for language classes that are closed under rational transductions and have
computable downward closures, separability by piecewise testable languages is
decidable.

As an abstraction, compared to the Parikh image (which counts the number of
occurrences of each letter), downward closures have the advantage of
guaranteeing regularity for any language. Most applications of Parikh
images, in contrast, require semilinearity, which fails for many
interesting language classes. An example of a class that lacks
semilinearity of Parikh images and thus spurred interest in computing
downward closures is that of the \emph{indexed languages}~\cite{Aho1968}
or, equivalently, those accepted by higher-order pushdown automata of
order~2~\cite{Maslov1976}.

It appears to be difficult to compute downward closures and there are few
language classes for which computability has been established.  Computability
is known for \emph{context-free languages} and \emph{algebraic
extensions}~\cite{Courcelle1991,vanLeeuwen1978}, \emph{0L-systems} and
\emph{context-free FIFO rewriting systems}~\cite{AbdullaBoassonBouajjani2001},
\emph{Petri net languages}~\cite{HabermehlMeyerWimmel2010}, and \emph{stacked
counter automata}~\cite{Zetzsche2015a}.  They are not computable for
reachability sets of \emph{lossy channel systems}~\cite{Mayr2003} and
\emph{Church-Rosser languages}~\cite{GruberHolzerKutrib2007}.  

This work presents a new general method for the computation of downward
closures.  It relies on a fairly simple idea and reduces the computation to the
so-called \emph{simultaneous unboundedness problem (SUP)}. The latter asks,
given a language $L\subseteq a_1^*\cdots a_n^*$, whether for each $k\in\N$,
there is a word $a_1^{x_1}\cdots a_n^{x_n}\in L$ such that $x_1,\ldots, x_n\ge
k$. This method yields new, sometimes greatly simplified, algorithms for each
of the computability results above. It also opens up a range of other language
classes to the computation of downward closures.

First, it implies computability for every language class that is closed under
rational transductions and exhibits effectively semilinear Parikh images.  This
re-proves computability for \emph{context-free languages} and \emph{stacked
counter automata}~\cite{Zetzsche2015a}, but also applies to many other classes,
such as the \emph{multiple context-free languages}~\cite{Seki1991}.  Second,
the method yields the computability for \emph{matrix
grammars}~\cite{DassowPaun1989,DassowPaunSalomaa1997}, a powerful grammar model
that generalizes Petri net and context-free languages.  Third, it is applied to
obtain computability of downward closures for the \emph{indexed languages}.

\section{Basic notions and results}
If $X$ is an alphabet, $X^*$ ($X^+$) denotes the set of (non-empty) words over
$X$.  The empty word is denoted by $\emptyWord\in X^*$.  For a symbol $x\in X$
and a word $w\in X^*$, let $|w|_x$ be the number of occurrences of $x$ in $w$.
If $w\in X^*$, we denote by $\alphabet(w)$ the set of symbols occurring in $w$.
For words $u,v\in X^*$, we write $u\preceq v$ if $u=u_1\cdots u_n$ and
$v=v_0u_1v_1\cdots u_nv_n$ for some $u_1,\ldots,u_n,v_0,\ldots,v_n\in X^*$.  It
is well-known that $\preceq$ is a well-quasi-order on $X^*$ and that therefore
the \emph{downward closure} 
\[ \Dclosure{L}=\{u\in X^* \mid \exists v\in L\colon u\preceq v\} \]
is regular for any $L\subseteq X^*$~\cite{Haines1969}.  If $X$ is
an alphabet, $X^\oplus$ denotes the set of maps $\alpha\colon X\to\N$, which
are called \emph{multisets}.  For $\alpha,\beta\in X^\oplus$, $k\in\N$ the
multisets $\alpha+\beta$ and $k\cdot\alpha$ are defined in the obvious way.  A
subset of $X^\oplus$ of the form 
\[ \{\mu_0+x_1\cdot \mu_1+\cdots+x_n\cdot \mu_n \mid x_1,\ldots,x_n\ge 0\} \] 
for $\mu_0,\ldots,\mu_n\in X^\oplus$ is called \emph{linear} and
$\mu_1,\ldots,\mu_n$ are its \emph{period elements}. A finite union of
linear sets is said to be \emph{semilinear}. The \emph{Parikh map} is the
map $\ParikhMap\colon X^*\to X^\oplus$ defined by $\Parikh{w}(x)=|w|_x$
for all $w\in X^*$ and $x\in X$. We lift $\ParikhMap$ to sets
in the usual way: $\Parikh{L}=\{\Parikh{w} \mid w\in L\}$. If
$\Parikh{L}=\Parikh{K}$, then $L$ and $K$ are said to be \emph{Parikh-equivalent}.

A \emph{finite automaton} is a tuple $(Q,X,E,q_0,F)$, where $Q$ is a finite set
of \emph{states}, $X$ is its input alphabet, $E\subseteq Q\times X^*\times Q$
is a finite set of \emph{edges}, $q_0\in Q$ is its \emph{initial state}, and
$F\subseteq Q$ is the set of its \emph{final states}. If there is a path
labeled $w\in X^*$ from state $p$ to $q$, we denote this fact by
$\autpath{p}{w}{q}$. The language accepted by $A$ is denoted $\Lang{A}$.

A \emph{(finite-state) transducer} is a tuple $(Q, X, Y, E, q_0, F)$, where
$Q$, $X$, $q_0$, $F$ are defined as for automata and $Y$ is its \emph{output
alphabet} and $E\subseteq Q\times X^*\times Y^*\times Q$ is the finite set of
its \emph{edges}. If there is a path from state $p$ to $q$ that reads the input
word $u\in X^*$ and outputs the word $v\in Y^*$, we denote this fact by
$\transpath{p}{u}{v}{q}$.  In slight abuse of terminology, we sometimes specify
transducers where an edge outputs a regular language instead of a word. 

For alphabets $X,Y$, a \emph{transduction} is a subset of $X^*\times
Y^*$. If $A$ is a transducer as above, then $\Trans{A}$ denotes
its generated \emph{transduction}, namely the set of all pairs $(u,v)\in
X^*\times Y^*$ such that $\transpath{q_0}{u}{v}{f}$ for some $f\in F$.
Transductions of the form $\Trans{A}$ are called \emph{rational}. For a
transduction $T\subseteq X^*\times Y^*$ and a language $L\subseteq X^*$,
we write $TL=\{v\in Y^* \mid \exists u\in L\colon (u,v)\in T\}$. A class
of languages $\C$ is called a \emph{full trio} if it is effectively
closed under rational transductions, i.e. if $TL\in\C$ for each $L\in\C$
and each rational transduction $T$.

Observe that for each full trio $\C$ and $L\in\C$, the language
$\Dclosure{L}$ is effectively contained in $\C$. By \emph{computing
downward closures} we mean finding a finite automaton for $\Dclosure{L}$
when given a representation of $L$ in $\C$. It will always be clear from
the definition of $\C$ how to represent languages in $\C$.

\subsection*{The simultaneous unboundedness problem} We come to the central
decision problem in this work.  Let $\C$ be a language class. The
\emph{simultaneous unboundedness problem} (SUP) \emph{for $\C$} is the
following decision problem:
\begin{description}
\item[Given] A language $L\subseteq a_1^*\cdots a_n^*$ in $\C$ for some alphabet $\{a_1,\ldots,a_n\}$.
\item[Question] Does $\Dclosure{L}$ equal $a_1^*\cdots a_n^*$?
\end{description}
The term ``simultaneous unboundedness problem'' reflects the fact that the
equality $\Dclosure{L}=a_1^*\cdots a_n^*$ holds if and only if for each
$k\in\N$, there is a word $a_1^{x_1}\cdots a_n^{x_n}\in L$ such that
$x_1,\ldots,x_n\ge k$. 

After obtaining the results of this work, the author learned that
\citeauthor{CzerwinskiMartens2014} considered a very similar decision
problem~\cite{CzerwinskiMartens2014}. Their \emph{diagonal problem} asks, given
a language $L\subseteq X^*$ whether for each $k\in\N$, there is a word $w\in L$
with $|w|_x\ge k$ for each $x\in X$. \citeauthor{CzerwinskiMartens2014} prove
that for full trios with a decidable diagonal problem, it is decidable whether
two given languages are separable by a piecewise testable language.  In fact,
their proof only requires decidability of the (ostensibly easier) SUP.  Here,
\Cref{dc} implies that in each full trio, the diagonal problem is decidable if
and only if the SUP is.

The following is the first main result of this work.
\begin{theorem}\label{dc}
Let $\C$ be a full trio. Then downward closures are computable for $\C$ if and
only if the SUP is decidable for $\C$.
\end{theorem}

The proof of \Cref{dc} uses the concept of simple regular expressions.  Let $X$
be an alphabet. An \emph{atomic expression} is a rational expression of the
form $(x\ratunion \emptyWord)$ with $x\in X$ or of the form $(x_1\ratunion
\cdots\ratunion x_n)^*$ with $x_1,\ldots,x_n\in X$.  A \emph{product} is a
(possibly empty) concatenation $a_1\cdots a_n$ of atomic expressions. A
\emph{simple regular expression (SRE)} is of the form $p_1\ratunion
\cdots\ratunion p_n$, where the $p_i$ are products. Given an SRE $r$, we write
$\Lang{r}$ for the language it describes.  

\Cref{dc} employs the following
result of \citeauthor{Jullien1969}~\cite{Jullien1969} (which was later rediscovered by
\citeauthor{Abdulla2004}~\cite{Abdulla2004}).
\begin{theorem}[\citeauthor{Jullien1969}~\cite{Jullien1969}]
Simple regular expressions describe precisely the downward closed languages.
\end{theorem}

We are now ready to prove \Cref{dc}.
\begin{proof}[Proof of \Cref{dc}]
Of course, if downward closures are computable for $\C$, then given a language
$L\subseteq a_1^*\cdots a_n^*$ in $\C$, we can compute a finite automaton for
$\Dclosure{L}$ and check whether $\Dclosure{L}=a_1^*\cdots a_n^*$. This proves
the ``only if'' direction.

For the other direction, we first observe that the emptiness problem can
be reduced to the SUP. Indeed, if $L\subseteq X^*$ and $T$ is the rational
transduction $X^*\times\{a\}^*$, then $TL\subseteq a^*$ and
$\Dclosure{(TL)}=a^*$ if and only if $L\ne\emptyset$.

Now, suppose the SUP is decidable for $\C$ and let $L\subseteq X^*$. Since we
know that $\Dclosure{L}$ is described by some SRE, we can enumerate SREs over
$X$ and are guaranteed that one of them will describe $\Dclosure{L}$.  Hence,
it suffices to show that given an SRE $r$, it is decidable whether
$\Lang{r}=\Dclosure{L}$. 

Since $\Lang{r}$ is a regular language, we can decide whether
$\Dclosure{L}\subseteq\Lang{r}$ by checking whether $\Dclosure{L}\cap
(X^*\setminus\Lang{r})=\emptyset$. This can be done because we can compute a
representation for $\Dclosure{L}\cap (X^*\setminus \Lang{r})$ in $\C$ and check
it for emptiness. It remains to be shown that it is decidable whether
$\Lang{r}\subseteq\Dclosure{L}$.

The set $\Lang{r}$ is a finite union of sets of the form
$\Dclosure{\{w_0\}}Y^*_1\Dclosure{\{w_1\}}\cdots Y^*_n\Dclosure{\{w_n\}}$ for
some $Y_i\subseteq X$, $Y_i\ne\emptyset$, $1\le i\le n$, and $w_i\in X^*$, $0\le
i\le n$. Therefore, it suffices to decide whether
$\Dclosure{\{w_0\}}Y_1^*\Dclosure{\{w_1\}}\cdots Y^*_n\Dclosure{\{w_n\}}
\subseteq \Dclosure{L}$.
Since $\Dclosure{L}$ is downward closed, this is
equivalent to
\begin{equation}
 w_0 Y_1^* w_1\cdots Y^*_n w_n\subseteq\Dclosure{L}.\label{dcinclusion}
\end{equation}
For each $i\in\{1,\ldots,n\}$, we define the word $u_i = y_1\cdots y_k$, where
$Y_i=\{y_1,\ldots,y_k\}$.  Observe that $w_0 Y_1^* w_1\cdots Y^*_n w_n\subseteq\Dclosure{L}$ holds if and
only if for every $k\ge 0$, there are numbers $x_1,\ldots,x_n\ge k$ such that
$w_0u_1^{x_1}w_1\cdots u_n^{x_n}w_n \in \Dclosure{L}$.  Moreover, if $T$ is the
rational transduction 
\[ T=\{(w_0u_1^{x_1}w_1\cdots u_n^{x_n}w_n,~a_1^{x_1}\cdots a_n^{x_n}) \mid x_1,\ldots,x_n\ge 0\}, \]
then 
$T(\Dclosure{L})=\{a_1^{x_1}\cdots a_n^{x_n} \mid w_0u_1^{x_1}w_1\cdots u_n^{x_n}w_n\in\Dclosure{L} \}$.
Thus, \cref{dcinclusion} is equivalent to
$\Dclosure{(T(\Dclosure{L}))}=a_1^*\cdots a_n^*$,  which is an instance of the SUP,
since we can compute a representation of $T(\Dclosure{L})$ in $\C$.
\end{proof}

Despite its simplicity, \Cref{dc} has far-reaching consequences for the
computability of downward closures. Let us record a few of them.
\begin{corollary}\label{dcpequi}
Suppose $\C$ and $\D$ are full trios such that given $L\in\C$, we can compute
a Parikh-equivalent $K\in\D$. If downward closures are computable for $\D$,
then they are computable for $\C$.
\end{corollary}
\begin{proof}
We show that the SUP is decidable for $\C$. Given $L\in\C$, $L\subseteq
a_1^*\cdots a_n^*$, we construct a Parikh-equivalent $K\in\D$.  Observe that
then $\Parikh{\Dclosure{K}}=\Parikh{\Dclosure{L}}$.  We compute a finite
automaton $A$ for $\Dclosure{K}$ and then a semilinear representation of
$\Parikh{\Lang{A}}=\Parikh{\Dclosure{K}}=\Parikh{\Dclosure{L}}$. Then
$\Dclosure{L}=a_1^*\cdots a_n^*$ if and only if some of the linear sets has for
each $1\le i\le n$ a period element containing $a_i$.  Hence, the SUP is
decidable for $\C$.
\end{proof}

Note that if a language class has effectively semilinear Parikh images, then we
can construct Parikh-equivalent regular languages. Therefore, the
following is a special case of \Cref{dcpequi}.
\begin{corollary}\label{dcsemilinear}
For each full trio with effectively semilinear Parikh images, downward closures
are computable.
\end{corollary}

\Cref{dcsemilinear}, in turn, provides computability of downward closures for a
variety of language classes. First, it re-proves the classical downward closure
result for \emph{context-free languages}~\cite{vanLeeuwen1978,Courcelle1991} and thus
\emph{algebraic extensions}~\cite{vanLeeuwen1978} (see~\cite{Zetzsche2015a} for a
simple reduction of the latter to the former).  Second, it yields a drastically
simplified proof of the computability of downward closures for \emph{stacked
counter automata}, which was shown in~\cite{Zetzsche2015a} using the machinery
of Parikh annotations. It should be noted, however, that the algorithm in
\cite{Zetzsche2015a} is easily seen to be primitive recursive, while this is
not clear for the brute-force approach presented here.  

\Cref{dcsemilinear} also implies computability of downward closures for
\emph{multiple context-free languages}~\cite{Seki1991}, which have received
considerable attention in computational linguistics.  As shown in
\cite{Seki1991}, the multiple context-free languages constitute a full trio and
exhibit effectively semilinear Parikh images.

Our next application of \Cref{dc} is an alternative proof of the computability
of downward closures for \emph{Petri net languages}, which was established by
\citeauthor{HabermehlMeyerWimmel2010}~\cite{HabermehlMeyerWimmel2010}.  Here,
by Petri net language, we mean sequences of transition labels of runs from an
initial to a final marking.
\citeauthor{CzerwinskiMartens2014}~\cite{CzerwinskiMartens2014} exhibit a
simple reduction of the diagonal problem for Petri net languages to the place
boundedness problem for Petri nets with one inhibitor arc, which was proven
decidable by
\citeauthor{BonnetFinkelLerouxZeitoun2012}~\cite{BonnetFinkelLerouxZeitoun2012}.
Since the Petri net languages are well-known to be a full
trio~\cite{Jantzen1979}, \Cref{dc} yields an alternative algorithm for downward
closures of Petri net languages.

We can also use \Cref{dcpequi} to extend the computability of downward closures
for Petri net languages to a larger class. \emph{Matrix grammars} are a
powerful formalism that is well-known in the area of regulated rewriting and
generalizes numerous other grammar
models~\cite{DassowPaun1989,DassowPaunSalomaa1997}.  They generate the
\emph{matrix languages}, a class which strictly includes both the context-free
languages and the Petri net languages. 
It is well-known that the matrix languages are a full trio and given a matrix grammar, one can construct a
Parikh-equivalent Petri net language~\cite{DassowPaunSalomaa1997}.
Thus, the following is a consequence of \Cref{dcpequi}.
\begin{corollary}
Downward closures are computable for matrix languages.
\end{corollary}

Finally, we apply \Cref{dc} to the \emph{indexed languages}. These were
introduced by \citeauthor{Aho1968}~\cite{Aho1968} and are precisely those
accepted by higher-order pushdown automata of order~2~\cite{Maslov1976}. Since
indexed languages do not have semilinear Parikh images, downward closures are a
promising alternative abstraction. 
\begin{theorem}\label{dcindexed}
Downward closures are computable for indexed languages.
\end{theorem}
The indexed languages constitute a full trio~\cite{Aho1968},
and hence the remainder of this work is devoted to showing that
their SUP is decidable. Note that since this class significantly
extends the 0L-languages~\cite{EhrenfeuchtRozenbergSkyum1976},
\Cref{dcindexed} generalizes the computability result of
\citeauthor{AbdullaBoassonBouajjani2001} for 0L-systems and context-free
FIFO rewriting systems~\cite{AbdullaBoassonBouajjani2001}.

\Cref{dcindexed} has an interesting consequence for computability of
downward closures in general. We will observe the following.
\begin{proposition}\label{undecidable}
Given an indexed language $L\subseteq a^*b^*$, it is undecidable whether
there is an $n\in\N$ with $a^nb^n\in L$.
\end{proposition}
First, this demonstrates that a slight variation of the SUP
is already undecidable. More importantly, \Cref{undecidable} means that
in automata that have access to a higher-order pushdown of order~2
and a very simple type of counter, reachability is undecidable: Given
a second-order pushdown automaton for $L$, one can use an additional
counter to accept $L\cap\{a^nb^n\mid n\ge 0\}$. Here, it even suffices
to use a blind counter (that is, one that can assume negative values and
has to be zero in accepting configurations~\cite{Greibach1978}) or a
reversal-bounded counter~\cite{Ibarra1978} (that is, one that switches
between incrementing and decrementing only a bounded number of times).

This is in contrast to the frequently used fact that \emph{semilinearity
is preserved by adding blind (or reversal bounded) counters}: When an
automata model guarantees effectively semilinear Parikh images, then the
model obtained by adding blind counters or reversal-bounded counters
still enjoys this property. Of course, this is not a precise statement,
but this fact has been discovered for various notions of storage
mechanisms~\cite{HarjuIbarraKarhumakiSalomaa2002, LohreySteinberg2008,
Zetzsche2013a}. Note that blind counters and reversal-bounded counters
are equivalent~\cite{Greibach1978} (see \cite{JantzenKurganskyy2003}
for a translation that is economic in the number of counters).
\Cref{dcindexed} and \Cref{undecidable} together imply that this
preservation has no analog for downward closures:
\begin{quote}
\emph{Adding blind (or reversal bounded) counters does not preserve
computability of downward closures.}
\end{quote}

\section{Indexed languages}
Let us define indexed grammars. The following definition is a slight
variation\footnote{We require that a nonterminal can only be replaced by a
terminal word if it has no index attached to it. It is easy to see that this
leads to the same languages~\cite{Smith2014}.} of the one
from~\cite{HopcroftUllman1979}.  An \emph{indexed grammar} is a tuple
$G=(N,T,I,P,S)$, where $N$, $T$, and $I$ are pairwise disjoint alphabets,
called the \emph{nonterminals}, \emph{terminals}, and \emph{index symbols},
respectively.  $P$ is the finite set of \emph{productions} of the forms $A\to
w$, $A\to Bf$, $Af\to w$, where $A,B\in N$, $f\in I$, and $w\in (N\cup T)^*$.
We regard a word $Af_1\cdots f_n$ with $f_1,\ldots,f_n\in I$ as a nonterminal
to which a stack is attached. Here, $f_1$ is the topmost symbol and $f_n$ is on
the bottom.  For $w\in (N\cup T)^*$ and $x\in I^*$, we denote by $[w,x]$ the
word obtained by replacing each $A\in N$ in $w$ by $Ax$. A word in $(NI^*\cup
T)^*$ is called a \emph{sentential form}. For $q,r\in (NI^*\cup T)^*$, we write
$q\grammarstep[G] r$ if there are words $q_1,q_2\in (NI^*\cup T)^*$, $A\in N$,
$p\in (N\cup T)^*$ and $x,y\in I^*$ such that $q=q_1 Ax q_2$, $r=q_1[p,y]q_2$,
and one of the following is true: \begin{enumerate}[label=(\roman*)]\item
$A\to p$ is in $P$, $p\in (N\cup T)^*\setminus T^*$, and $y=x$, \item $A\to p$
is in $P$, $p\in T^*$, and $y=x=\emptyWord$, \item $A\to pf$ is in $P$ and
$y=fx$, or \item $Af\to p$ is in $P$ and $x=fy$. \end{enumerate} The language
\emph{generated by $G$} is 
$\Lang{G}=\{ w\in T^* \mid S\grammarsteps[G] w\}$,
where $\grammarsteps[G]$ denotes the reflexive transitive closure of
$\grammarstep[G]$.

We will often assume that our indexed grammars are in \emph{normal form}, which means
that every production is in one of the following forms:
 \begin{align*}
 &\text{(i) $A\to Bf$},		&\text{(ii) $Af\to B$},		& &\text{(iii) $A\to uBv$},	& &\text{(iv) $A\to BC$},	& &\text{(v) $A\to w$},
 \end{align*}
with $A,B,C\in N$, $f\in I$, and $u,v,w\in T^*$.
Productions of these forms are called \emph{push}, \emph{pop}, \emph{output},
\emph{split}, and \emph{terminal} productions, respectively. The normal form
can be attained just like the Chomsky normal form of context-free
grammars.

\begin{example}\label{examplegrammar}
Let $G=(N,T,I,P,S)$ be the indexed grammar with $N=\{S,T,A,B\}$,
$T=\{a,b\}$, $I=\{f,g\}$, and the productions
\begin{align*}
S  & \to Sf, & S  & \to Sg, & S & \to UU, & U & \to\varepsilon, \\
Uf & \to A,  & Ug & \to B,  & A & \to Ua, & B & \to Ub.
\end{align*}
Then it is easy to see that $\Lang{G}=\{ww \mid w\in\{a,b\}^*\}$.
\end{example}

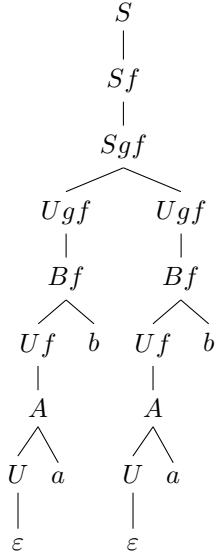
\begin{figure}
\begin{center}
\begin{tikzpicture}
\tikzset{level distance=25pt}
\Tree [ .{$S$} 
        [ .{$Sf$}
          [ .{$Sgf$}
            [ .{$Ugf$}
              [ .{$Bf$}
                [ .{$Uf$}
                  [ .{$A$}
                    [ .{$U$}
                      [ .{$\varepsilon$} ] ]
                    [ .{$a$} ] ] ]
		[ .{$b$} ] ] ]
            [ .{$Ugf$}
              [ .{$Bf$}
                [ .{$Uf$}
                  [ .{$A$}
                    [ .{$U$}
                      [ .{$\varepsilon$} ] ]
                    [ .{$a$} ] ] ]
		[ .{$b$} ] ] ] ] ] ]
\end{tikzpicture}
\end{center}
\caption{Derivation tree for the grammar in \Cref{examplegrammar} with yield $abab$.}\label{derivationtree}
\end{figure}

\emph{Derivation trees} are always unranked trees with labels in
$NI^*\cup T\cup \{\emptyWord\}$ and a very straightforward analog to
those of context-free grammars. If $t$ is a labeled tree, then its
\emph{yield}, denoted $\yield{t}$, is the word spelled by the labels of
its leaves. For an example for the grammar from \Cref{examplegrammar},
see \Cref{derivationtree}.

\subsection*{Overview} The SUP for indexed grammars does not
seem to easily reduce to a decidable problem. In the case $L\subseteq a^*$, the
SUP is just the finiteness problem, for which \citeauthor{Hayashi1973}
presented a procedure using his pumping lemma~\cite{Hayashi1973}. 
However, neither Hayashi's nor any of the other pumping or shrinking
lemmas~\cite{Smith2014,Gilman1996b,Kartzow2011,Parys2012} appears
to yield decidability of the SUP. Therefore, this work employs a
different approach: Given an indexed grammar $G$ with $\Lang{G}\subseteq
a_1^*\cdots a_n^*$, we apply a series of transformations, each
preserving the simultaneous unboundedness (\cref{step:interval,step:productive,step:partitioned,step:transducers,step:breadthbounded}). These
transformations leave us with an indexed grammar in which the number
of nonterminals appearing in sentential forms is bounded. This allows
us to construct an equivalent finite-index scattered context grammar
(\cref{step:semilinearity}), a type of grammars that is known to exhibit
effectively semilinear Parikh images.

\subsection*{An undecidability result} Before proving decidability of
the SUP for indexed languages, we prove \Cref{undecidable}.

\begin{proof}[Proof of \Cref{undecidable}]
We provide a reduction from the \emph{Post correspondence problem
(PCP)}, which asks, given an alphabet $X$ and morphisms
$\alpha,\beta\colon X^*\to\{1,2\}^*$, whether there is a $w\in X^+$
with $\alpha(w)=\beta(w)$. It is well-known that this problem is
undecidable~\cite{Post1946}.

For a word $w\in \{1,2\}^*$, let $\nu(w)\in\N$ be the number obtained
by interpreting the word $w$ as a $2$-adic representation.
This means, for $w\in \{1,2\}^*$, we have
\begin{align*} \nu(\emptyWord)=0, && \nu(w1)=2\cdot \nu(w)+1, &&  \nu(w2)=2\cdot \nu(w)+2.  \end{align*}
Given an alphabet $X$ and two morphisms $\alpha,\beta\colon
X^*\to\{1,2\}^*$, we shall construct an indexed grammar $G$ with
\begin{equation}
 \Lang{G}=\{a^{\nu(\alpha(w))}b^{\nu(\beta(w))} \mid w\in X^+ \}. \label{undeclang}
\end{equation}
Note that this establishes the \lcnamecref{undecidable}: Since
the map $\nu\colon\{1,2\}^*\to\N$ is injective, the equation
\labelcref{undeclang} implies that $\Lang{G}\cap\{a^nb^n\mid
n\ge 0\}\ne\emptyset$ if and only if there is a $w\in X^+$ with
$\alpha(w)=\beta(w)$.

For the sake of simplicity of the other proofs, our definition of
indexed grammars restricts the syntax of productions. To make the
the description of $G$ more convenient, we allow one more case as a
shorthand: In the following, when we write $Ax_0\to Bx_1\cdots x_n$
for nonterminals $A,B$ and index symbols $x_0,\ldots,x_n$, then this
represents $n+2$ productions, $Ax_0\to Z_n$, $Z_i\to Z_{i-1}x_i$ for
$1\le i\le n$, and $Z_0\to B$, where $Z_0,\ldots,Z_n$ are nonterminals
occurring nowhere else.

The grammar $G$ has nonterminals $S,U,A,\bar{A},B,\bar{B}$ (and those
resulting from using shorthands), index symbols $I=X\cup \{1,2\}$ (we
assume that $X\cap\{1,2\}=\emptyset$), and terminals $T=\{a,b\}$. The
first set of productions allows us to derive all sentential forms $AwBw$
for $w\in X^+$. For each $x\in X$, we have:
\begin{align*}
	S\to Ux, && U\to Ux, && U\to AB.
\end{align*}
We also have productions that allow the nonterminals $A,\bar{A}$
($B$,$\bar{B}$) to replace an index symbol $x$ with $\alpha(x)$
($\beta(x)$):
\begin{align}
	Cx & \to C\alpha(x) &  & \text{for each $x\in X$ and $C\in\{A,\bar{A}\}$}, \label{prodalpha}\\
	Cx & \to C\beta(x)  &  & \text{for each $x\in X$ and $C\in\{B,\bar{B}\}$}. \label{prodbeta}
\end{align}
These guarantee that for every $w\in X^*$ and $C\in
\{A,\bar{A}\}$, the sentential forms $Cw$ and $C\alpha(w)$ derive
the same terminal words. Analogously, for $w\in X^*$ and $C\in
\{B,\bar{B}\}$, the sentential forms $Cw$ and $C\beta(w)$ derive the
same terminal words.

Together with \labelcref{prodalpha,prodbeta}, the next set
of productions turns the sentential form $Aw$ and $Bw$ into
$a^{\nu(\alpha(w))}$ and $b^{\nu(\beta(w))}$, respectively: For each
$C\in \{A,B\}$, we have
\begin{align}
	C1\to C\bar{C}, && C2\to C\bar{C}\bar{C}, && \bar{C}d\to \bar{C}\bar{C}.\label{proddyadic}
\end{align}
Finally, to obtain terminal words, we have the productions 
\begin{align}
	A&\to \emptyWord & \bar{A}&\to a, & B&\to \emptyWord, & \bar{B}&\to b.\label{prodterminal}
\end{align}
It remains to be shown that our grammar meets the goal in
\cref{undeclang}. Because of the productions \labelcref{prodalpha,prodbeta},
it suffices to show that for $w\in\{1,2\}^*$, the sentential form $Aw$
($Bw$) derives precisely one terminal word, namely $a^{\nu(\alpha)}$
($b^{\nu(\beta(w))}$). For symmetry reasons, we only show that $Aw$
derives precisely the word $a^{\nu(\alpha(w))}$.

We proceed by induction and strengthen the statement slightly.
Namely, we claim that for $w_0,\ldots,w_n\in \{1,2\}^*$, the
sentential form $Aw_0\bar{A}w_1\cdots\bar{A}w_n$ derives precisely
the word $a^{\nu(w_0)+m}$, where $m=\sum_{i=1}^n 2^{|w_i|}$.
We use noetherian induction with respect to the set of finite
sequences of natural numbers, ordered lexicographically, and to
the words $w_0,\ldots,w_n\in\{1,2\}^*$, we assign the sequence
$(|w_0|,\ldots,|w_n|)$. Now the induction step is just the observation
that for every derivation step $Aw_0\bar{A}w_1\cdots \bar{A}w_n\grammarstep
Az_0\bar{A}z_1\cdots\bar{A}z_k$ with $w_0,\ldots,w_n,z_0,\ldots,z_k\in
\{1,2\}^*$, we have $(|z_0|,\ldots,|z_k|) < (|w_0|,\ldots,|w_n|)$ in the
lexicographical order and also
\[ \nu(w_0)+\sum_{i=1}^n 2^{|w_i|}=\nu(z_0)+\sum_{i=1}^k 2^{|z_i|}. \]
This can be seen by inspecting the productions \labelcref{proddyadic}
and noticing that for all words $w\in \{1,2\}^*$, we have
\begin{align*}
	\nu(1w)=2^{|w|}+\nu(w), && \nu(2w)=2\cdot 2^{|w|}+\nu(w).
\end{align*}
Moreover, the claim is true in the case $w_0=\cdots=w_n=\emptyWord$.
Indeed, the sentential form $A\bar{A}^n$ can only derive the word $a^n$
and $n=\nu(\emptyWord)+\sum_{i=1}^n 2^0$.
\end{proof}

\subsection{Triple construction}\label{tripleconstruction} We begin
with the triple construction, a standard tool in the theory of grammars
that we will use on several occasions. Suppose we have an indexed
grammar $G=(N,T,I,P,S)$ in normal form and a finite-state transducer
$A=(Q,T,X,E,q_0,F)$. The triple construction is usually employed to
prove closure under rational transductions, i.e. to build a grammar
$G_A=(N_A, T, I, P_A, S_A)$ such that $\Lang{G_A}=V\Lang{G}$, where
$V=\Trans{A}$. More precisely, $N_A$ consists of all triples $(p,B,q)$
with $p,q\in Q$ and $B\in N$ and they satisfy:
\begin{align} (p,B,q)x\grammarsteps[G_A] y \iffSpace\text{if and only if}\iffSpace \exists z\in T^*\colon Bx\grammarsteps[G] z,~~\transpath{p}{z}{y}{q}.\label{tripleequivalence} \end{align}
For the construction, we assume that the edges in $A$ are all of the
form $(p,t,\emptyWord,q)$ or $(p,\emptyWord,x,q)$ for $p,q\in Q$, $t\in
T$, $x\in X$. Furthermore, we assume that $A$ has only one final state.
Suppose $p,q\in Q$, $B\in N$, and consider the languages
\[ L_{p,B,q}=\{v\in X^* \mid \exists B\to u\in P,~u\in T^*,~\transpath{p}{u}{v}{q}~\text{in $A$} \}. \]
Since each of these sets is regular, we can construct an automaton $U$ with state
set $\bar{Q}$ such that for each $p,q\in Q$ and $B\in N$, there are states
$r_{p,B,q},s_{p,B,q}\in\bar{Q}$ with
\[ \autpath{r_{p,B,q}}{w}{s_{p,B,q}}~\text{in $U$} \iffSpace\text{if and only if} \iffSpace w\in L_{p,B,q} \]
for $w\in X^*$.  We are now ready to describe the grammar $G_A$. Its set of
nonterminals is $N_A=(Q\times N\times Q)\cup (\bar{Q}\times\bar{Q})$.  The
first type of productions are the following.  For $r,s\in\bar{Q}$ and each edge
$\autpath{r}{x}{t}$ in $U$, we have a production $(r,s)\to x(t,s)$.
Furthermore, for each $s\in\bar{Q}$, we have the production
$(s,s)\to\emptyWord$. Since these will be the only productions with left-hand
side in $\bar{Q}\times\bar{Q}$, we will have
\[ (r,s)\grammarsteps[G_A] w \iffSpace\text{if and only if}\iffSpace\autpath{r}{w}{s}~\text{in $U$} \]
for $r,s\in\bar{Q}$ and $w\in X^*$. For $p,q,r\in Q$, $B,C,D\in N$, $g\in I$, $u,v\in T^*$, $G_A$ has productions
\begin{align*}
(p,B,q) & \to (p, C, q)g && \text{for each $B\to Cg\in P$}, \\
(p,B,q)g & \to (p, C, q) && \text{for each $Bg\to C\in P$}, \\
(p,B,q)&\to u(p', C, q')v && \text{for each $B\to uCv\in P$, $p',q'\in Q$} \\
       &                  && \text{with $\autpath{p}{u}{p'}$, $\autpath{q'}{v}{q}$ in $A$} \\
(p,B,q) & \to (p, C, r)(r, D, q) && \text{for each $B\to CD\in P$}.
\end{align*}
Moreover, it has the production $(p,B,q)\to (r_{p,B,q}, s_{p,B,q})$ for each
$p,q\in Q$ and $B\in N$. Now it is easy to verify that \cref{tripleequivalence}
is satisfied.  Therefore, we let $(q_0, S, f)$ be the start symbol of $G_A$,
where $q_0$ and $f$ are the initial and the final state, respectively, of $A$.
Then, in particular, we have $\Lang{G_A}=V\Lang{G}$, where $V=\Trans{A}$.

\subsection{Regular index sets} 
We now analyze the structure of index words that facilitate certain
derivations. Let $G=(N,T,P,S)$ be an indexed grammar, $A\in N$
a nonterminal and $R\subseteq T^*$ a regular language. We write
$\IW{G}{A}{R}$ for the set of index words that allow $A$ the derivation
of a word from $R$. This means
\[ \IW{G}{A}{R}=\{x\in I^* \mid \exists y\in R\colon Ax\grammarsteps[G] y \}.  \]
The following lemma is essentially equivalent to the fact that the set
of stack contents from which an alternating pushdown system can reach
a final configuration is regular~\cite{BouajjaniEsparzaMaler1997}. We
include here a proof in the terminology of indexed grammars.
\begin{lemma}\label{regularindices}
For an indexed grammar $G$, a nonterminal $A$, and a regular language $R$, the
language $\IW{G}{A}{R}$ is effectively regular.
\end{lemma}
\begin{proof}
Let $G=(N,T,I,P,S)$.  First of all, we may assume that $G$ is in normal form,
since bringing an indexed grammar in normal form does not affect the languages
$\IW{G}{A}{R}$. Suppose $A'$ is a transducer such that for some states $p,q$,
we have $\transpath{p}{z}{y}{q}$ in $A'$ if and only if $z=y$ and $z\in R$.
In particular, $\IW{G}{A}{R}$ equals $\IW{G_{A'}}{(p,A,q)}{T^*}$, where $G_{A'}$ is
obtained using the triple construction.  Therefore, it means no loss of
generality to assume that $R=T^*$.  Hence, we may discard the generated
terminals and assume that in $G$, every production is of the form $B\to Cf$,
$Bf\to C$, or $B\to w$ with $w\in N^*$.

\newcommand{\cf}{\mathsf{cf}}
Our next step is to construct a grammar $G'$ with the same nonterminal and
index symbols as $G$ such that (i) $\IW{G'}{A}{R}=\IW{G}{A}{R}$ and (ii) if
$Aw\grammarsteps[G']\emptyWord$, then $\emptyWord$ can be derived from $Aw$
without using push productions. We do this by successively computing grammars
$G_i=(N,T,I,P_i,S)$ for $i\in\N$ such that $P_0\subseteq P_1\subseteq\cdots$.
We initialize $P_0=P$. Suppose $G_i$ is already defined and that every
productions in $P_i$ is of the form $B\to Cf$, $Bf\to C$, or $B\to w$ for some
$w\in N^*$.  In the following, we say that a production is a \emph{nonterminal
production} if it is of the form $B\to w$ with $B\in N$ and $w\in N^+$.
Consider the language
\[ K^{(i)}_B = \{ w\in N^+ \mid B\grammarstepsp[G_i] w\}, \]
where $\grammarstepp[G_i]$ is the restricted derivation relation that only
permits nonterminal productions. Then $K^{(i)}_B$ is clearly context-free.
Hence, we know that also the language $L^{(i)}_{B,f}=V_fK^{(i)}_B$ is
context-free, where $V_f$ is the rational transduction that on input
$w=B_1\cdots B_k$, outputs all words $C_1\cdots C_k$ for which there are
productions $B_jf\to C_j$ in $P_i$ for $1\le j\le k$.  Observe that
$L^{(i)}_{B,f}$ consists of all those sentential forms of $G_i$ reachable from
$Bf$, $B\in N$, $f\in I$, by first applying only nonterminal productions and
then applying at each position a production that pops $f$: Since
$\grammarstepp[G_i]$ only allows nonterminal productions, the production
sequence for $w\in K^{(i)}_B$ is applicable to $Bf$ as well. (Recall that our
definition of indexed grammars forbids the application of terminal productions
to nonterminals with a non-empty index.)

The context-freeness of $L^{(i)}_{B,f}$ allows us to compute the set
$W^{(i)}_{B,f}\subseteq \Powerset{N}$ with
\[ W^{(i)}_{B,f} = \{ \alphabet(w) \mid w\in L^{(i)}_{B,f} \}. \]
The set $W^{(i)}_{B,f}$ describes all combinations of nonterminals that
can result when applying to $Bf$ a number of nonterminal productions and
then at each position a production popping $f$. We are ready to describe
the productions in $P_{i+1}$. For each subset $X\subseteq N$, we pick
a word $w_X\in N^*$ with $\alphabet(w_X)=X$ and $|w_X|=|X|$. We obtain
$P_{i+1}$ by adding to $P_i$ the production $C\to w_X$ for each $C\in N$
and $X\in W^{(i)}_{B,f}$ such that $C\to Bf\in P_i$ with $B\in N$, $f\in
I$.

Note that since we only add productions, we have $\IW{G_i}{A}{R}\subseteq
\IW{G_{i+1}}{A}{R}$ and the construction guarantees
$\IW{G_i}{A}{R}=\IW{G_{i+1}}{A}{R}$: Since the added $w_X$ contains all the
nonterminals of the corresponding word in $L^{(i)}_{B,f}$, a derivation of
$\emptyWord$ in $G_{i+1}$ can easily be turned into a derivation in $G_i$ by
replicating subtrees in the derivation tree.

Since we only add productions of the form $C\to w$ with $|w|\le
|N|$, there must come an $i$ with $P_{i+1}=P_i$. This means that for
each $C\in N$ and each $u\in L^{(i)}_{B,f}$ such that $C\to Bf\in
P_i$, we have some production $C\to w$ with $\alphabet(w)=\alphabet(u)$.
Therefore, in $G_i$, for $A\in N$, $v\in I^*$, the sentential form
$Av$ can derive $\emptyWord$ if and only if it can do so without using
push productions: For each derivation of $\emptyWord$ from $Av$ that
uses a push production, we can bypass this push production and all
corresponding pop productions by using one of the added productions
$C\to w_X$. Thus, a derivation of $\emptyWord$ with a minimal number of
occurrences of push productions has to avoid them altogether.

This allows us to construct a finite automaton for
$\reverse{\IW{G_i}{A}{R}}$. Here, for a language $U\subseteq X^*$,
$\reverse{U}$ denotes the set of words from $U$ in reverse. As the
automaton reads index words from right to left, it maintains the set
of nonterminals $B$ for which the currently read suffix $v$ satisfies
$Bv\grammarsteps[G_i] \emptyWord$. The set of states of our automaton is
therefore the power set of $N$ and its initial state is 
\[ q_0 = \{ B\in N \mid B\grammarstepspp[G_i] \emptyWord \}, \]
where $\grammarsteppp[G_i]$ is the restricted derivation relation that
only permits productions with a left-hand side in $N$ and a right-hand
side in $N^*$. As transitions, the automaton has for every $X\subseteq N$ 
and $f\in I$ an edge
\[ \autpath{X}{\quad f\quad}{\{ B\in N \mid W^{(i)}_{B,f}\subseteq X\}} \]
Note that we can again compute the initial state of the automaton using
context-freeness arguments. The final states of the automaton are all
those sets $X\subseteq N$ that contain $A$. Then, the automaton clearly
accepts $\reverse{\IW{G_i}{A}{R}}=\reverse{\IW{G}{A}{R}}$.
\end{proof}

\subsection{Interval grammars}\label{step:interval}
We want to make sure that each nonterminal can only derive words in some fixed
`interval' $a_i^*\cdots a_j^*$.  An \emph{interval grammar} is an indexed
grammar $G=(N,T,I,P,S)$ in normal form together with a map $\iota\colon N\to
\N\times\N$, called \emph{interval map}, such that for each $A\in N$ with
$\iota(A)=(i,j)$, we have
\begin{enumerate}[label=(\roman*)]\item $1\le i\le j\le n$, \item\label{interval:derivable} if
$Ax\grammarsteps[G] u$ for $x\in I^*$ and $u\in T^*$, then $u\in a_i^*\cdots
a_j^*$, and \item if $S\grammarsteps[G] u\,Ax\,v\,By\,w$ with $u,v,w\in
(NI^*\cup T)^*$, $B\in N$, $x,y\in I^*$, and $\iota(B)=(k,\ell)$, then $j\le
k$.\end{enumerate}

\begin{proposition}\label{interval}
For each indexed grammar $G$ with $\Lang{G}\subseteq a_1^*\cdots a_n^*$, there
is an equivalent interval grammar.
\end{proposition}
\begin{proof}
Let $G=(N,T,I,P,S)$ be an indexed grammar in normal form such that
$\Lang{G}\subseteq a_1^*\cdots a_n^*$. Our new grammar has nonterminals 
\[ N'=\{(i,A,j) \mid A\in N,~~1\le i\le j\le n \} \]
and productions
\begin{align*}
(i,A,j)f&\to (i,B,j)		&& \text{for each $Af\to B\in P$}, \\ 
(i,A,j)&\to (i,B,j)f		&& \text{for each $A\to Bf\in P$}, \\
(i,A,j)&\to u(r,B,s)v		&& \text{for each $A\to uBv\in P$, $i\le r\le s\le j$}, \\
        &                       && \text{$u\in a_i^*\cdots a_r^*$ and $v\in a_s^*\cdots a_j^*$}, \\
(i,A,j)&\to (i,B,k)(k,C,j)	&& \text{for each $A\to BC\in P$ and $i\le k\le j$}, \\
(i,A,j)&\to w			&& \text{for each $A\to w\in P$ with $w\in a_i^*\cdots a_j^*$}
\end{align*}
where $A,B,C\in N$ and $f\in I$.  As the new start symbol, we choose $(1,S,n)$.
Then, setting $\iota((i,A,j))=(i,j)$ for each $A\in N$ and $i,j\in\N$ clearly
yields an interval grammar and its equivalence to $G$ is easily verified.
\end{proof}
\subsection{Productive grammars}\label{step:productive}
We will also need our grammar to be `productive', meaning that 
every derivable sentential form and every nonterminal in it 
contribute to the derived terminal words.
A production is called \emph{erasing} if its right-hand side is the empty word.
A grammar is \emph{non-erasing} if it contains no erasing productions.
Moreover, a word $u\in (NI^*\cup T)^*$ is \emph{productive} if there is some
$v\in T^*$ with $u\grammarsteps[G] v$.  We call an indexed grammar $G$
\emph{productive} if \begin{enumerate*}[label=(\roman*)]\item it is non-erasing
and \item whenever $u\in (NI^*\cup T)^*$ is productive and $u\grammarsteps[G]
u'$ for $u'\in (NI^*\cup T)^*$, then $u'$ is productive as
well.\end{enumerate*} The following \lcnamecref{intervalproductive} is shown in
two steps.  First, we construct an interval grammar and then use
\Cref{regularindices} to encode information about the current index word in
each nonterminal. This information is then used, among other things, to prevent the
application of productions that lead to non-productive sentential forms.
The \lcnamecref{intervalproductive} clearly implies that the SUP for indexed grammars can
be reduced to the case of productive interval grammars.
\begin{proposition}\label{intervalproductive}
For each indexed grammar $G$ with $\Lang{G}\subseteq a_1^*\cdots a_n^*$, one
can construct a productive interval grammar $G'$ with
$\Lang{G'}=\Lang{G}\setminus\{\emptyWord\}$.
\end{proposition}

The proof of \Cref{intervalproductive} relies on a construction that is used
again for the proof of \Cref{productivetransduction}, so we describe it for
general indexed grammars. 

\newcommand{\image}[1]{\mathsf{im}\,#1}
\newcommand{\domain}[1]{\mathsf{dom}\,#1}
\todo{explain what we are doing now}
By \Cref{regularindices}, the languages $\IW{G}{A}{T^+}$ and $\IW{G}{A}{\{\emptyWord\}}$
are effectively regular. This means, we can construct a deterministic finite
automaton that reads a word over $I$ in reverse and, after reading the suffix $u\in I^*$, 
maintains in its state the set of all nonterminals $A$ with $u\in\IW{G}{A}{T^+}$
as well as the set of those $A$ for which $u\in\IW{G}{A}{\{\emptyWord\}}$.

Let us formalize this. There is a finite set $Q$, an element $q_0\in Q$, maps
$\sigma_0,\sigma_+\colon Q\to N$, and a map $\cdot\colon I\times Q\to Q$ such
that if we extend the latter map to $\cdot\colon I^*\times Q\to Q$ via $ua\cdot
q=u\cdot(a\cdot q)$ and $\emptyWord \cdot q=q$ for $a\in I$, $u\in I^*$, $q\in
Q$, then 
\begin{align*} 
\sigma_0(u\cdot q_0)&=\{ A\in N \mid Au\grammarsteps[G]\emptyWord \}, \\   
\sigma_+(u\cdot q_0)&=\{ A\in N \mid \exists v\in T^+\colon Au\grammarsteps[G]v \} 
\end{align*}
for each $u\in I^*$.

The idea behind the construction of $\hat{G}$ is to encode into each nonterminal the
state in $Q$ reached by reading its current index. Hence, as nonterminals, we
have the set $\hat{N}=N\times Q$.  In order to be able to update this state, we also
need to encode such states into the index words themselves. Here, each index symbol
will encode the state reached by reading the suffix to its right. Thus, the
index symbols in $\hat{G}$ are $\hat{I}=I\times Q$. Formally, we want to achieve the following.
Let $g\colon (NI^*\cup T)^*\to (\hat{N}\hat{I}^*\cup T)^*$ be the function with
\[ g(Af_n\cdots f_1)=(A, q_n)(f_n,q_{n-1})\cdots (f_1,q_0), \]
for $f_n,\ldots,f_1\in I$, where $q_i=f_i\cdots f_1q_0$ for $1\le i\le n$ and
\[ g(u_0A_1w_1u_1\cdots A_mw_mu_m)=u_0g(A_1w_1)u_1\cdots g(A_mw_m)u_m \]
for $u_0,\ldots,u_m\in T^*$, $A_1,\ldots,A_m\in N$, $w_1,\ldots,w_m\in I^*$. Then, we want
the grammar $\hat{G}$ to satisfy
\begin{align} Aw\grammarsteps[G] v \iffSpace\text{if and only if}\iffSpace g(Aw) \grammarsteps[\hat{G}] v \label{productivegoal} \end{align}
for $A\in N$, $w\in I^*$, and $v\in T^+$. $\hat{G}$ has the productions
\begin{align*}
(A,q)(f,q')&\to (B,q')		&& \text{for each $Af\to B\in P$  with $B\in\sigma_+(q')$, $q=f\cdot q'$}, \\
(A,q)&\to (B, f\cdot q)(f,q)	&& \text{for each $A\to Bf\in P$  with $B\in\sigma_+(f\cdot q)$}, \\
(A,q)&\to u(B,q)v		&& \text{for each $A\to uBv\in P$ with $B\in\sigma_+(q)$}, \\
(A,q)&\to (B,q)			&& \text{for each $A\to BC\in P$  with $B\in\sigma_+(q)$, $C\in\sigma_0(q)$}, \\
(A,q)&\to (C,q)			&& \text{for each $A\to BC\in P$  with $B\in\sigma_0(q)$, $C\in\sigma_+(q)$}, \\
(A,q)&\to (B,q)(C,q)		&& \text{for each $A\to BC\in P$  with $B\in\sigma_+(q)$, $C\in\sigma_+(q)$}, \\
(A,q_0)&\to u			&& \text{for each $A\to u\in P$   with $u\in T^+$}.
\end{align*}
Furthermore, the start symbol of $\hat{G}$ is $(S,q_0)=g(S)$.  Each of the directions of
\cref{productivegoal} now follows by induction on the number of derivation
steps. Hence, we have $\Lang{\hat{G}}=\Lang{G}\setminus\{\emptyWord\}$. Let us prove
that $\hat{G}$ is productive.  Suppose the partial function $h\colon (NI^*\cup T)^* \to
(\hat{N}\hat{I}^*\cup T)^*$ is defined as the restriction of $g$ to those words
\[ x=u_0A_1w_1u_1\cdots A_mw_mu_m, \]
(with $u_0,\ldots,u_m\in T^*$, $A_1,\ldots,A_m\in N$, $w_1,\ldots,w_m\in
I^*$) where for each index $i\in\{1,\ldots,m\}$, the set $\sigma_+(w_i\cdot q_0)$
contains $A_i$. Then it follows from \cref{productivegoal} that every
$u\in\image{h}$ is productive. Furthermore, by induction on $n$, one
can show that $u\grammarstepsn[\hat{G}]{n} v$, $v\in T^+$, implies
$u\in\image{h}$. Thus, $u$ is productive in $\hat{G}$ if and only
if $u\in\image{h}$. Moreover, an inspection of the productions in
$\hat{G}$ reveals that if $u\grammarstep[\hat{G}]v$ and $u\in\image{h}$,
then $v\in\image{h}$. Thus, if $u\in (\hat{N}\hat{I}^*\cup T)^*$ is
productive, then every sentential form reachable from $u$ is productive.
Hence, $\hat{G}$ is productive.

\begin{proof}[Proof of \Cref{intervalproductive}]
Using \Cref{interval}, we construct an interval grammar $H$ with
$\Lang{H}=\Lang{G}$.  Let $\hat{H}=(\hat{N},T,\hat{I},\hat{P},\hat{S})$ be
obtained from $H$ as above.  Then $\hat{H}$ is productive and generates
$\Lang{\hat{H}}=\Lang{H}\setminus\{\emptyWord\}$.  We define $\hat{\iota}\colon
\hat{N}\to\N\times\N$ by $\hat{\iota}((A,q))=\iota(A)$. Then $\hat{H}$,
together with $\hat{\iota}$, is clearly a productive interval grammar with
$\Lang{\hat{H}}=\Lang{G}\setminus\{\emptyWord\}$.
\end{proof}

\subsection{Partitioned grammars}\label{step:partitioned}
Our next step is based on the following observation. Roughly speaking, in an
interval grammar, in order to generate an unbounded number of $a_i$'s, there
have to be derivation trees that contain either
\begin{enumerate}[label=(\roman*)]\item an unbounded number of incomparable
(with respect to the subtree ordering) $a_i$-subtrees (i.e.  subtrees with
yield in $a_i^*$) or \item a bounded number of such subtrees that themselves
have arbitrarily large yields.\end{enumerate}
In a partitioned grammar, we designate for each $a_i$, whether we allow
arbitrarily many $a_i$-subtrees (each of which then only contains a
single $a_i$) or we allow exactly one $a_i$-subtree (which is then
permitted to be arbitrarily large). The symbols of the former kind will be dubbed
`direct'.

Let us formalize this.  A nonterminal $A$ in an interval grammar is called
\emph{unary} if $\iota(A)=(i,i)$ for some $1\le i\le n$. A \emph{partitioned}
grammar is an interval grammar $G=(N,T,I,P,S)$, with interval map $\iota\colon
N\to\N\times\N$, together with a subset $D\subseteq T$ of \emph{direct symbols}
such that for each $a_i\in T$, the following holds:
\begin{enumerate*}[label=(\roman*)]
\item If $a_i\in D$, then there is no $A\in N$ with $\iota(A)=(i,i)$, and
\item if $a_i\notin D$ and $t$ is a derivation tree of $G$, then all
occurrences of $a_i$ are contained in a single subtree whose root contains a
unary nonterminal.
\end{enumerate*}
In other words, direct symbols are never produced through unary nonterminals,
but always directly through non-unary ones. If, on the other hand, $a_i$ is not
direct, then all occurrences of $a_i$ stem from one occurrence of a suitable
unary symbol.
The next \lcnamecref{partitioned} clearly reduces the SUP for indexed grammars 
to the case of partitioned grammars.
\begin{proposition}\label{partitioned}
Let $G$ be a productive interval grammar with $\Lang{G}\subseteq a_1^*\cdots
a_n^*$.  Then, one can construct partitioned grammars $G_1,\ldots,G_m$ such
that $\Dclosure{\Lang{G}}$ equals $a_1^*\cdots a_n^*$ if and only if
$\Dclosure{\Lang{G_i}}=a_1^*\cdots a_n^*$ for some $1\le i\le m$.
\end{proposition}
\begin{proof}
Suppose $G$ is a productive interval grammar.  We prove the
\lcnamecref{partitioned} by constructing for each subset $D\subseteq T$ a
partitioned grammar $G_D$ and then show that $\Dclosure{\Lang{G}}=a_1^*\cdots
a_n^*$ if and only if $\Dclosure{\Lang{G_D}}=a_1^*\cdots a_n^*$ for some
$D\subseteq T$.  Observe that for $n=1$, $G$ is already a partitioned grammar
with $D=\emptyset$. Therefore, we may assume that $n\ge 2$.

Since $G$ is a productive interval grammar, we may assume that each of its
productions is in one of the following forms:
\begin{enumerate}[label=(\roman*)]
\item $A\to Bf$ with $\iota(A)=(i,j)$, $\iota(B)=(r,s)$ and $i\le r\le s\le j$,
\item $Af\to B$ with $\iota(A)=(i,j)$, $\iota(B)=(r,s)$ and $i\le r\le s\le j$,
\item $A\to uBv$ with $\iota(A)=(i,j)$, $\iota(B)=(r,s)$, $u\in a_i^*\cdots a_r^*$, $v\in a_s^*\cdots a_j^*$,
\item $A\to BC$ with $\iota(A)=(i,j)$, $\iota(B)=(p,q)$, $\iota(C)=(r,s)$ and $i\le j\le p\le q\le r\le s\le j$,
\item $A\to u$ with $\iota(A)=(i,j)$ and $u\in a_i^*\cdots a_j^*$
\end{enumerate}
with $A,B,C\in N$. A production that is not of this form that is used in a
derivation would allow the grammar to violate condition
\cref{interval:derivable} of interval grammars. Hence, every production that is
not in one of these forms can be safely removed. By introducing new intermediate
nonterminals, we can therefore even assume that every production is in one the
following forms:
\begin{enumerate}[label=(\roman*)]
\item $A\to Bf$ with $\iota(A)=\iota(B)$,
\item $Af\to B$ with $\iota(A)=\iota(B)$,
\item $A\to uBv$ with $\iota(A)=(i,j)$, $\iota(B)=(r,s)$, $u\in a_i^*\cdots a_r^*$, $v\in a_s^*\cdots a_j^*$,
\item $A\to BC$ with $\iota(A)=(i,j)$, $\iota(B)=(p,q)$, $\iota(C)=(r,s)$ and $i\le j\le p\le q\le r\le s\le j$,
\item $A\to u$ with $\iota(A)=(i,j)$ and $u\in a_i^*\cdots a_j^*$
\end{enumerate}
Suppose $D\subseteq T$. First, we construct the grammar $G'_D$ from $G$. Here,
the essential idea is to replace each maximal subtree whose root has a label
$Ax$ with $A\in N$, $x\in I^*$, $\iota(A)=(i,i)$ and $a_i\in D$ by a single
node labeled $a_i$.  The resulting trees are the derivation trees of $G'_D$,
which then has no nonterminals $A$ with $\iota(A)=(i,i)$ and $a_i\in D$.

Because of our normal form, whenever a unary nonterminal is introduced in $G$
that does not already stem from a nonterminal with the same $\iota$-value, the
left-hand side of the production is a non-unary nonterminal.  Hence, consider a
production $A\to w$ in $G$ such that $\iota(A)=(i,j)$ with $i<j$ and $w\in
(N\cup T)^*$. Let $w'\in (N\cup T)^*$ be obtained from $w$ by replacing each
$B\in N$, $\iota(B)=(k,k)$, $a_k\in D$, with the symbol $a_k$.  

\begin{enumerate}[label=(\roman*)]
\item If $|w'|_N\ge 1$, we add the production $A\to w'$. Note that then,
$A\to w'$ can be applied whenever $A\to w$ is applied (Recall that
productions with a right-hand side in $T^*$ can only be applied when the
index word is empty).
\item If $w'\in T^*$, the production $A\to w'$ is not applicable when
the $A$ in the sentential form still carries a non-empty index word. In
this case, we introduce a fresh nonterminal $E$, set $\iota(E)=(i,j)$,
and add productions $A\to E$, $Ef\to E$ for each $f\in I$, and $E\to
w'$. Then, whenever $A\to w$ is applied, we can instead apply $A\to E$,
then remove the index with $Ef\to E$, and finally apply $E\to w'$.
\end{enumerate}
Moreover, we remove all nonterminals $A$ with $\iota(A)=(i,i)$,
$a_i\in D$ and all productions containing such nonterminals 

Now in fact, the derivation trees of $G'_D$ are precisely those obtained
from derivations trees $t$ of $G$ by replacing every maximal subtree
whose root is labeled $Ax$, $A\in N$, $x\in I^*$, $\iota(A)=(i,i)$,
$a_i\in D$, with a node labeled $a_i$ and, if necessary, adding a path
of productions $Ef\to E$.

Note that since $G$ is productive and $\iota(A)=(i,i)$, every
word derivable from $A$ (together with an index word) is contained in $a_i^+$.
Furthermore, every occurrence of $A$ in a derivable sentential form of $G$ is
also able to derive a word in $a_i^+$. This means, for each $u\in\Lang{G'_D}$,
there is a word $v\in \Lang{G}$ with $u\subword v$.  Hence, we have
$\Dclosure{\Lang{G'_D}}\subseteq\Dclosure{\Lang{G}}$.

Consider a derivation tree of $G'_D$ or of $G$. We call a node \emph{$i$-node}
if its label is $a_i$ or some $A\in N$ with $\iota(A)=(i,i)$. If, in addition,
the $i$-node has no $i$-node as an ancestor, it is an \emph{$i$-root}. A subtree whose root node is an $i$-root
of the derivation tree is called \emph{$i$-subtree}.

As a second step, we construct $G_D$ from $G'_D$ so that the following holds:
The derivation trees of $G_D$ are precisely those obtained from derivation
trees of $G'_D$ by essentially deleting for each $a_i\in T\setminus D$ all but
one $i$-subtree (`essentially' because we have to rename the remaining
nonterminals). Of course, if the deletion of subtrees leaves behind a leaf
labeled with a nonterminal, we attach an $\emptyWord$-labeled node below it.
The construction of $G_D$ is achieved by letting each nonterminal carry a
function $\alpha\colon T\setminus D\to \{0,1,\omega\}$.  Here, $\alpha(a_i)=1$
indicates that the one allowed $i$-subtree is somewhere below the current node;
$\alpha(a_i)=0$ means that the $i$-subtree is located elsewhere in the
derivation tree; and $\alpha(a_i)=\omega$ indicates that the current node is
part of the $i$-subtree. In particular, the new start symbol carries the
function $\alpha$ with $\alpha(a_i)=1$ every $a_i\in T\setminus D$. It is easy
to adjust the productions to use and update these functions $\alpha$. 

Now, every word in $\Lang{G_D}$ is obtained from a word in $\Lang{G'_D}$ by deleting
for each $a_i\in T\setminus D$ the yields of all but one $i$-subtree. Hence,
for each $u\in\Lang{G_D}$, there is a $v\in\Lang{G'_D}$ with $u\subword v$.
Thus, we have
$\Dclosure{\Lang{G_D}}\subseteq\Dclosure{\Lang{G'_D}}\subseteq\Dclosure{\Lang{G}}$.
This means, if $\Dclosure{\Lang{G_D}}=a_1^*\cdots a_n^*$, then
$\Dclosure{\Lang{G}}=a_1^*\cdots a_n^*$.  It remains to be shown that if
$\Dclosure{\Lang{G}}=a_1^*\cdots a_n^*$, then there is some $D\subseteq T$ with
$\Dclosure{\Lang{G_D}}=a_1^*\cdots a_n^*$.

Suppose $\Dclosure{\Lang{G}}=a_1^*\cdots a_n^*$. Then there is a sequence
$t_1,t_2,\ldots$ of derivation trees of $G$ such that $a_1^k\cdots
a_n^k\subword \yield{t_k}$.  For each derivation tree $t$, let $\sigma_i(t)$ be
the number of $i$-subtrees in $t$.  By Dickson's Lemma, we can
pick a subsequence $t'_1,t'_2,\ldots$ of $t_1,t_2,\ldots$ such that for $1\le
i\le n$, $\sigma_i$ is monotonically increasing on $t'_1,t'_2,\ldots$.  We
claim that with
\[ D=\{a_i\in T \mid \text{$\sigma_i$ is unbounded on $t'_1,t'_2,\ldots$}\}, \]
the grammar $G_D$ satisfies $\Dclosure{\Lang{G_D}}=a_1^*\cdots a_n^*$.  By
definition of $D$, we can find a subsequence $t''_1,t''_2,\ldots$ of
$t'_1,t'_2,\ldots$ such that $\sigma_i(t''_k)\ge k$ for every $a_i\in D$.  Let
$s_k=\overline{t''_k}$ be the derivation tree of $G'_D$ corresponding to
$t''_k$ of $G$ as above.  Then we have $a_i^k\subword a_i^{\sigma_i(t''_k)}\subword
\yield{s_k}$ for $a_i\in D$.  Since we only change $i$-subtrees for $a_i\in D$
when going from $t''_k$ to $s_k$, we still have $a_i^k\subword \yield{s_k}$ for
$a_i\in T\setminus D$ and thus $a_1^k\cdots a_n^k\subword \yield{s_k}$.

The choice of $D$ guarantees that $\sigma_i$ is bounded on
$s_1,s_2,\ldots$ for every $a_i\in T\setminus D$. Hence, there is an
$\ell\in\N$ with $\sigma_i(s_k)\le \ell$ for every $k\in\N$ and $a_i\in
T\setminus D$.  This means, if $\tau_i(t)$ is the maximal length of a yield of
an $i$-subtree of $t$, then $\tau_i$ is unbounded on $s_1,s_2,\ldots$ for each
$a_i\in T\setminus D$. Indeed, if $\tau_i$ were bounded on $s_1,s_2,\ldots$ by
$B\in\N$, then $\yield{s_k}$ would contain at most $\ell\cdot B$ occurrences of
$a_i$ for $a_i\in T\setminus D$, contradicting $a_i^k\subword \yield{s_k}$. We
can therefore find a subsequence $s'_1,s'_2,\ldots$ of $s_1,s_2,\ldots$ such
that $\tau_i(s'_k)\ge k$ for $a_i\in T\setminus D$. Note that since this is a
subsequence of $s_1,s_2,\ldots$, it automatically satisfies
$a_i^k\subword\yield{s'_k}$ for $a_i\in D$.

Let us now turn the trees $s'_1,s'_2,\ldots$ into derivation trees
$r_1,r_2,\ldots$ of $G_D$. We do this by deleting, for each $a_i\in T\setminus
D$, from $s'_k$ all $i$-subtrees but the one with the longest yield (and
renaming the remaining nonterminals to obtain derivation trees of $G_D$).
Again, if this deletion leaves behind a leaf labeled by a nonterminal, we
attach an $\emptyWord$-labeled node beneath it.  Clearly, each $r_k$ is a
derivation tree of $G_D$.  Observe that $a_1^k\cdots a_n^k\subword\yield{r_k}$.
Indeed, if $a_i\in D$, then $a_i^k\subword \yield{s'_k}$ and thus
$a_i^k\subword \yield{r_k}$. If $a_i\in T\setminus D$, then $\tau_i(s'_k)\ge k$
and hence $a_i^k\subword\yield{r_k}$.  Therefore, we have
$\Dclosure{\Lang{G_D}}=a_1^*\cdots a_n^*$.
\end{proof}

\subsection{Constructing transducers}\label{step:transducers}
The last step in our proof (\cref{step:semilinearity}) will be to solve
the SUP in the case where we have a bound on the number of nonterminals
in reachable sentential forms. The only obstacle to such a bound are the
unary nonterminals corresponding to terminals $a_i\notin D$: All other
nonterminals have $\iota(A)=(i,j)$ with $i<j$ and there can be at most
$n-1$ such symbols in a sentential form. However, for each $a_i\notin
D$, there is at most one subtree with a corresponding unary nonterminal
at its root. Our strategy is therefore to replace these problematic
subtrees so as to bound the nonterminals: Instead of unfolding the
subtree generated from $u\in NI^*$, we apply a transducer to $u$.

\newcommand{\Ninf}{\N\cup\{\infty\}}
In order to guarantee that the replacement does not affect whether
$\Dclosure{\Lang{G}}$ equals $a_1^*\cdots a_n^*$, we employ a slight
variant\footnote{The difference is that we have an equivalence on
partial instead of total functions.} of the equivalence that gives
rise to the \emph{cost functions} of Colcombet~\cite{Colcombet2013}.  If
$f\colon X\to\Ninf$ is a partial function, we say that $f$ is \emph{unbounded
on $E\subseteq X$} if for each $k\in\N$, there is some $x\in E$ with $f(x)\ge
k$ (in particular, $f(x)$ is defined).  If $g\colon X\to\Ninf$ is another
partial function, we write $f\approx g$ if for each subset $E\subseteq X$, we
have: $f$ is unbounded on $E$ if and only if $g$ is unbounded on $E$. Note that
if $h\colon Y\to X$ is a partial function and $f\approx g$, then $h\circ
f\approx h\circ g$. Now, we compare the transducer and the original grammar
on the basis of the following partial functions. Given an indexed grammar $G=(N,T,I,P,S)$
and a transducer $A$ with $\Trans{A}\subseteq NI^*\times T^*$, we define
the partial functions $f_G,f_A\colon NI^*\to \Ninf$ by
\begin{align*}
 f_G(u)&=\sup \{ |v| \mid v\in T^*,~u\grammarsteps[G] v \}, \\
 f_A(u)&=\sup \{ |v| \mid v\in T^*,~(u,v)\in \Trans{A} \}.
\end{align*}
Note that here, $\sup M$ is undefined if $M$ is the empty set.
\begin{proposition}\label{transduction}
Given an indexed grammar $G$, one can construct a finite-state transducer $A$
such that $f_A\approx f_G$.
\end{proposition}

\subsection*{Productivity} In the proof of \Cref{transduction}, we assume that
$G$ is productive. The following \lcnamecref{productivetransduction} justifies
this.  A partial function $h\colon X^*\to Y^*$ is called \emph{rational} if 
\[ \{(u,v)\in X^*\times Y^* \mid f(u)=v\} \]
is a rational transduction.
\begin{lemma}\label{productivetransduction}
Given an indexed grammar $G$, one can construct a productive grammar $G'$ and a
rational partial function $h$ such that $f_G\approx h\circ f_{G'}$. 
\end{lemma}
\begin{proof}
Consider the grammar $\hat{G}$ and the partial function $h$ constructed in the
proof of \Cref{intervalproductive}. Since $\hat{G}$ is productive and $h$ is
clearly rational, it suffices to show that $h\circ f_{\hat{G}}\approx f_G$.

Note that $h$ is defined on $u\in NI^*$ if and only if there is some $v\in T^+$
with $u\grammarsteps[G] v$. This means, $u\in\domain{h}$ if and only if
$f_G(u)\ge 1$. Furthermore, if $u\in\domain{h}$, then \cref{productivegoal}
implies that $f_{\hat{G}}(h(u))=f_G(u)$. Hence $h\circ f_{\hat{G}}$ and $f_G$
agree on $\domain{h}$ and are both bounded on $NI^*\setminus\domain{h}$. This
clearly implies $h\circ f_{\hat{G}}\approx f_G$.
\end{proof}

Now it suffices indeed to prove \Cref{transduction} for productive grammars: If
we can construct a finite-state transducer $A$ with $f_A\approx f_{G'}$ and
$A'$ is the transducer that first computes $h$ and then applies $A$, we have
$f_{A'}=h\circ f_A\approx h\circ f_{G'}\approx f_G$.  Hence, we assume that $G$
is productive. 

The construction of the transducer will
involve deciding the \emph{finiteness problem} for indexed languages, which
asks, given $G$, whether $\Lang{G}$ is finite. Its decidability has been shown
by Rounds~\cite{Rounds1970} (and later again by Hayashi~\cite[Corollary
5.1]{Hayashi1973}). 

\begin{theorem}[\citeauthor{Rounds1970}~\cite{Rounds1970}]\label{finiteness}
The finiteness problem for indexed languages is decidable.
\end{theorem}

Let $R=\{Bw \mid B\in N,~w\in I^*,~w\in\IW{G}{B}{T^*}\}$. Then $f_G$ is clearly
undefined on words outside of $R$. Therefore, it suffices to exhibit a
finite-state transducer $A$ with $f_A|_R\approx f_G$: The regularity of $R$
means we can construct a transducer $A'$ with $f_{A'}=f_A|_R$.  
In order to
prove the relation $f_A|_R\approx f_G$, we employ the concept of shortcut
trees.

\subsection*{Shortcut trees} Note that since $G$ is productive, the label
$\emptyWord$ does not occur in derivation trees for $G$. Let $t$ be such a
derivation tree. Let us inductively define the set of \emph{shortcut trees} for
$t$.  Suppose $t$'s root $r$ has the label $\ell\in NI^*\cup T$.  If $\ell\in
N\cup T$, then the only shortcut tree for $t$ consists of just one node with
label $\ell$.  If $\ell=Bfv$, $B\in N$, $f\in I$, $v\in I^*$, then the shortcut
trees for $t$ are obtained as follows. We choose a set $U$ of nodes in $t$ such that
\begin{conditions}
\item each path from $r$ to a leaf contains precisely one node in $U$,
\item the label of each $x\in U$ either equals $Cv$ for some $C\in N$ or belongs to $T$,
\item each node on the path from $r$ to any $x\in U$ has a label of the form $Cuv$ with $C\in N$ and $u\in I^*$.
\end{conditions}
For each such choice of $U=\{x_1,\ldots,x_n\}$, we take shortcut trees
$t_1,\ldots,t_n$ for the subtrees of $x_1,\ldots,x_n$ and create a new shortcut
tree for $t$ by attaching $t_1,\ldots,t_n$ to a fresh root node.  The root node
carries the label $B$.  This is how all shortcut trees for $t$ are obtained.
For an example of a shortcut tree for a derivation tree, see \Cref{exampleshortcuttree}.
Note that every shortcut tree for $t$ has height $|\ell|-1$. We also 
call these \emph{shortcut trees from $\ell$}.

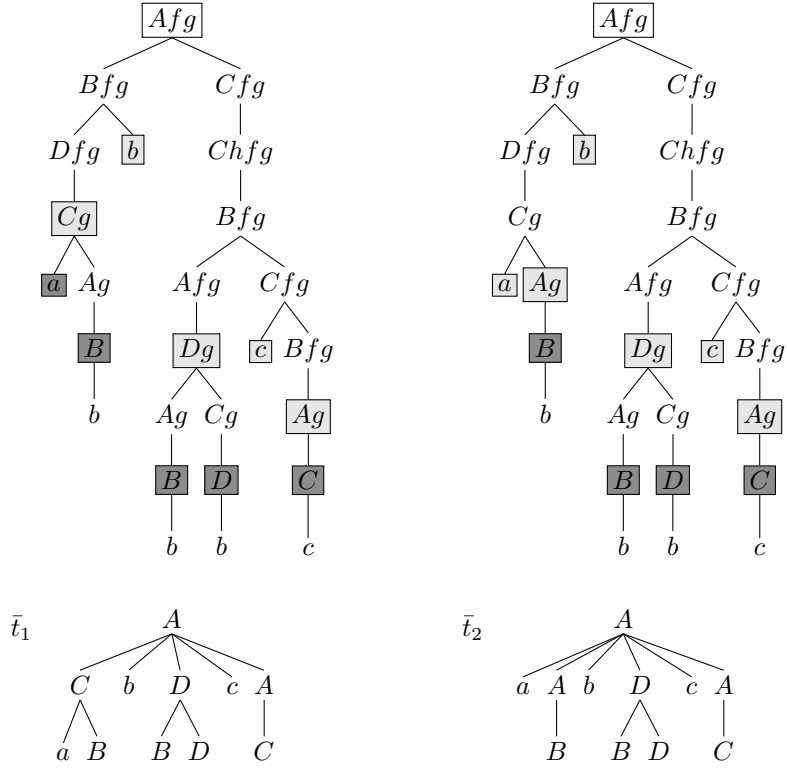
\begin{figure}
\begin{tikzpicture}[
   level distance=25pt,
   inner sep=2pt,
   shortcut1/.style={rectangle,draw=black,fill=gray!0},
   shortcut2/.style={rectangle,draw=black,fill=gray!20},
   shortcut3/.style={rectangle,draw=black,fill=gray!90}]
\Tree [ .\node[shortcut1](r1){$Afg$};
        [ .{$Bfg$}
          [ .{$Dfg$}
            [ .\node[shortcut2](r21){$Cg$};
              [ .\node[shortcut3](r31){$a$};
              ]
              [ .{$Ag$} 
                [ .\node[shortcut3](r32){$B$};
                  [ .{$b$} 
                  ]
                ]
              ]
            ]
          ]
          [ .\node[shortcut2](r22){$b$};
          ]
        ]
        [ .{$Cfg$}
          [ .{$Chfg$}
            [ .{$Bfg$}
              [ .{$Afg$} 
                [ .\node[shortcut2](r23){$Dg$};
                  [ .{$Ag$}
                    [ .\node[shortcut3](r33){$B$};
                      [ .{$b$}
                      ]
                    ]
                  ]
                  [ .{$Cg$}
                    [ .\node[shortcut3](r34){$D$};
                      [ .{$b$} 
                      ]
                    ]
                  ]
                ]
              ]
              [ .{$Cfg$} 
                [ .\node[shortcut2](r24){$c$};
                ]
                [ .{$Bfg$}
                  [ .\node[shortcut2](r25){$Ag$};
                    [ .\node[shortcut3](r35){$C$};
                      [ .{$c$}
                      ]
                    ]
                  ]
                ]
              ]
            ]
          ]
        ]
      ]

\begin{scope}[level distance=25pt, shift={(0cm,-8cm)}]
\Tree [ .\node(s1){$A$};
        [ .\node(s21){$C$};
	  [ .\node(s31){$a$};
	  ]
	  [ .\node(s32){$B$};
	  ]
	]
	[ .\node(s22){$b$};
	]
	[ .\node(s23){$D$};
	  [ .\node(s33){$B$};
	  ]
	  [ .\node(s34){$D$};
	  ]
	]
	[ .\node(s24){$c$};
	]
	[ .\node(s25){$A$};
	  [ .\node(s35){$C$};
	  ]
	]
      ]
\node at (-2cm,0) {$\bar{t}_1$};
\end{scope}
\begin{scope}[shift={(6cm, 0cm)}]
\Tree [ .\node[shortcut1](r1){$Afg$};
        [ .{$Bfg$}
          [ .{$Dfg$}
            [ .{$Cg$}
              [ .\node[shortcut2](r21){$a$};
              ]
              [ .\node[shortcut2](r22){$Ag$};
                [ .\node[shortcut3](r32){$B$};
                  [ .{$b$} 
                  ]
                ]
              ]
            ]
          ]
          [ .\node[shortcut2](r23){$b$};
          ]
        ]
        [ .{$Cfg$}
          [ .{$Chfg$}
            [ .{$Bfg$}
              [ .{$Afg$} 
                [ .\node[shortcut2](r24){$Dg$};
                  [ .{$Ag$}
                    [ .\node[shortcut3](r33){$B$};
                      [ .{$b$}
                      ]
                    ]
                  ]
                  [ .{$Cg$}
                    [ .\node[shortcut3](r34){$D$};
                      [ .{$b$} 
                      ]
                    ]
                  ]
                ]
              ]
              [ .{$Cfg$} 
                [ .\node[shortcut2](r25){$c$};
                ]
                [ .{$Bfg$}
                  [ .\node[shortcut2](r26){$Ag$};
                    [ .\node[shortcut3](r35){$C$};
                      [ .{$c$}
                      ]
                    ]
                  ]
                ]
              ]
            ]
          ]
        ]
      ]

\end{scope}
\begin{scope}[level distance=25pt, shift={(6cm,-8cm)}]
\Tree [ .\node(s1){$A$};
        [ .\node(s21){$a$};
	]
	[ .\node(s22){$A$};
	  [ .\node(s32){$B$};
	  ]
	]
	[ .\node(s23){$b$};
	]
	[ .\node(s24){$D$};
	  [ .\node(s33){$B$};
	  ]
	  [ .\node(s34){$D$};
	  ]
	]
	[ .\node(s25){$c$};
	]
	[ .\node(s26){$A$};
	  [ .\node(s35){$C$};
	  ]
	]
      ]
\node at (-2cm,0) {$\bar{t}_2$};
\end{scope}
\end{tikzpicture}
\caption{Example of a derivation tree and two possible shortcut trees.
The trees above are two drawings of the same derivation tree. In each
of the derivation trees, the boxed nodes induce nodes in the shortcut
tree below it. The shading of each boxed node indicates the level of
its corresponding node in the shortcut tree. Here, $A,B,C,D$ are
nonterminals, $f,g,h$ are index symbols, and $a,b,c$ are terminals. }
\label{exampleshortcuttree}
\end{figure}

\newcommand{\children}[1]{\mathsf{w}(#1)}
\newcommand{\rootNode}[1]{\mathsf{root}(#1)}
In other words, a shortcut tree is obtained by successively choosing a
sentential form such that the topmost index symbol is removed, but the
rest of the index is not touched. For example, the chosen sentential
forms in \Cref{exampleshortcuttree} are $Afg$, $CgbDgcAg$, and $aBbDcC$
on the left-hand side and $Afg$, $aAgbDgcAg$, and $aBbBDcC$ on the
right-hand side. Note that if $\bar{t}$ is a shortcut tree for a
derivation tree $t$, then we have $|\yield{\bar{t}}|\le|\yield{t}|$. On
the other hand, every derivation tree has a shortcut tree with the same
yield. Thus, if we define $\bar{f}_G\colon NI^*\to\Ninf$ by
\[ \bar{f}_G(u)=\sup\{ |\yield{\bar{t}}| \mid \text{$\bar{t}$ is a shortcut tree from $u$} \} \]
then we clearly have $\bar{f}_G\approx f_G$. Therefore, in order to prove
$f_A|_R\approx f_G$, it suffices to show $f_A|_R\approx \bar{f}_G$.
Let us describe the transducer $A$.  For $B,C\in N$ and $g\in
I$, consider the language
$L_{B,g,C} = \{ w\in (N\cup T)^* \mid Bg \grammarstepsp[G] w,~|w|_C\ge 1\}$.
Here, $\grammarstepp[G]$ denotes the restricted derivation relation that
forbids terminal productions.
Then $L_{B,g,C}$ is the set of words $\children{\rootNode{\bar{t}}}$ for
shortcut trees $\bar{t}$ of derivation trees from $Bg$ (or, equivalently, $Bgv$
with $v\in I^*$) such that $C$ occurs in $\children{\rootNode{\bar{t}}}$. Here, 
$\rootNode{\bar{t}}$ denotes the root node of $\bar{t}$ and $\children{\rootNode{\bar{t}}}$ is the word
consisting of the labels of the root's child nodes. Each $L_{B,g,C}$ belongs to the class of indexed languages, which is
a full trio and has a decidable finiteness
and emptiness problem. Hence, we can compute the following function, which will
describe $A$'s output. Pick an $a\in T$ and define for each $B,C\in N$ and $g\in I$:
\newcommand{\Output}[3]{\mathsf{Out}(#1,#2,#3)}
\[ \Output{B}{g}{C}=\begin{cases} \{a\}^* & \text{if $L_{B,g,C}$ is infinite}, \\
\{a\} & \text{if $L_{B,g,C}$ is finite and $L_{B,g,C}\cap (N\cup T)^{\ge 2}\ne\emptyset$}, \\
\{\emptyWord\} & \text{if $L_{B,g,C}\ne\emptyset$ and $L_{B,g,C}\subseteq N\cup T$}, \\
\emptyset & \text{if $L_{B,g,C}=\emptyset$}.
\end{cases} \]
Note that for each $B,C\in N$ and $g\in I$, precisely one of the conditions on
the right holds.  The transducer $A$ has states $\{q_0\}\cup N$ and edges
$\transedge{q_0}{B}{\{\emptyWord\}}{B}$ and
$\transedge{B}{g}{\Output{B}{g}{C}}{C}$
for each $B,C\in N$ and $g\in I$. $A$'s initial state is $q_0$ and its final
states are all those $B\in N$ with $B\grammarsteps[G] w$ for some $w\in T^*$.
Hence, the runs of $A$ on a word $Bw\in R$ correspond to paths (from
root to leaf) in shortcut trees from $Bw$.  Here, the productivity of the words
in $R$ guarantees that every run of $A$ with input from $R$ does in
fact arise from a shortcut tree in this way.

Suppose $A$ performs a run on input $Bw\in R$, $|w|=k$, and produces
the outputs $a^{n_1},\ldots,a^{n_k}$ in its $k$ steps that read $w$.  Then the
definition of $\Output{\cdot}{\cdot}{\cdot}$ guarantees that there is a
shortcut tree $\bar{t}$ such that the run corresponds to a path in
which the $i$-th node has at least $n_i+1$ children. In particular, $\bar{t}$ has at
least $n_1+\cdots+n_k$ leaves. Therefore, we have $f_A(Bw)\le\bar{f}_G(Bw)$.

It remains to be shown that if $\bar{f}_G$ is unbounded on $E\subseteq R$, then
$f_A$ is unbounded on $E$. For a
tree $t$, let $\delta(t)$ denote the maximal number of children of any node and
let $\beta(t)$ denote the maximal number of branching nodes (i.e. those with at
least two children) on any path from root to leaf. We use the following simple combinatorial fact, for which we do not provide a proof.
\begin{lemma}\label{treesunbounded}
In a set of trees, the number of leaves is unbounded if and only if $\delta$ is
unbounded or $\beta$ is unbounded.
\end{lemma}
Suppose $\bar{f}_G$ is unbounded on $E\subseteq R$. Then there is a sequence of
shortcut trees $t_1,t_2,\ldots$ from words in $E$ such that
$|\yield{t_1}|,|\yield{t_2}|,\ldots$ is unbounded. This means
$\delta$ or $\beta$ is unbounded on
$t_1,t_2,\ldots$. Note that if $t$ is a shortcut tree from $Bw\in R$, then 
the path in $t$ with $\beta(t)$ branching nodes gives rise to a run of
$A$ on $Bw$ that outputs at least $\beta(t)$ symbols. Hence, $f_A(Bw)\ge\beta(t)$.
Thus, if $\beta$ is unbounded on $t_1,t_2,\ldots$, then $f_A$ is unbounded on $E$.

Suppose $\delta$ is unbounded on $t_1,t_2,\ldots$. Let $x$ be an inner node of
a shortcut tree $\bar{t}$.  Then the subtree of $x$ is also a shortcut tree,
say of a derivation tree $t$ from $Bgw\in R$ with $B\in N$, $g\in I$, $w\in
I^*$. Moreover, $x$ has a child node with a label $C\in N$ (otherwise, it would
be a leaf of $\bar{t}$). We say that $(B,g,C)$ is a \emph{type} of $x$ (note
that a node may have multiple types). Since $\delta$ is unbounded on
$t_1,t_2,\ldots$ and there are only finitely many possible types, we can pick a type
$(B,g,C)$ and a subsequence $t'_1,t'_2,\ldots$ such that each $t'_k$ has an
inner node $x_k$ with at least $k$ children and type $(B,g,C)$. This means
there are nodes of type $(B,g,C)$ with arbitrarily large numbers of children
and hence $L_{B,g,C}$ is infinite. We can therefore choose any $t'_i$ and a run
of $A$ that corresponds to a path involving $x_i$. Since $L_{B,g,C}$ is
infinite, this run outputs $\{a\}^*$ in the step corresponding to $x_i$.
Moreover, this run reads a word in $E$ and hence $f_A$ is unbounded on $E$.
This proves $f_A|_R\approx \bar{f}_G$ and thus \Cref{transduction}.

\subsection{Breadth-bounded grammars}\label{step:breadthbounded}
A \emph{breadth-bounded grammar} is an indexed grammar, together with
a bound $k\in\N$, such that each of its reachable sentential forms
contains at most $k$ nonterminals. \Cref{transduction} allows us to
prove the following.
\begin{proposition}\label{partitionedbreadthbounded}
Let $G$ be a partitioned grammar with $\Lang{G}\subseteq a_1^*\cdots a_n^*$.
Then, one can construct a breadth-bounded grammar $G'$ with $\Lang{G'}\subseteq
a_1^*\cdots a_n^*$ such that $\Dclosure{\Lang{G}}=a_1^*\cdots a_n^*$ if and
only if $\Dclosure{\Lang{G'}}=a_1^*\cdots a_n^*$.
\end{proposition}
The proof comprises two steps. First, we build a breadth-bounded
grammar that, instead of unfolding the derivation trees below
unary nonterminals, outputs their index words as terminal words,
which results in a breadth-bounded grammar. Then, we apply our
transducer from \Cref{transduction} to the resulting subwords.
Since the breadth-bounded grammars generate a full trio, the
\lcnamecref{partitionedbreadthbounded} follows. The former is a matter
of inspecting the triple construction.
\begin{lemma}\label{breadthboundedfulltrio}
The languages generated by breadth-bounded grammars form a full trio.
\end{lemma}
\begin{proof}
Let $G$ be a breadth-bounded grammar and $A$ be a
finite-state transducer with $V=\Trans{A}$. In order to prove the
\lcnamecref{breadthboundedfulltrio}, we need to exhibit a
breadth-bounded grammar that generates $V\Lang{G}$. Consider
the grammar $G_A$ resulting from the triple construction (see
\cref{tripleconstruction}).

Since $G_A$ generates $V\Lang{G}$, it suffices to show that $G_A$
is breadth-bounded. This, however, follows directly from the
breadth-boundedness of $G$: Every sentential form of $G_A$ is obtained
from a sentential form of $G$ by replacing nonterminals $B\in N$
by symbols $(p,B,q)$ with $p,q\in Q$ or by symbols $(r,s)$ with
$r,s\in\bar{Q}$. Hence, if in $G$, every sentential form contains at
most $k$ nonterminals, this is also true of $G_A$.
\end{proof}

Let $G=(N,T,I,P,S)$ be a partitioned grammar with direct symbols
$D\subseteq T$. We will be interested in derivations where the unary
nonterminals are not rewritten. Therefore, we have the derivation
relation $\grammarstep[G,D]$, in which $u\grammarstep[G,D] v$ if and
only if $u \grammarstep[G] v$ and the employed production does not
replace a unary nonterminal. This allows us to define
\[ \PLang{G} = \{ w\in (UI^*\cup D)^* \mid S\grammarsteps[G,D] w \}, \]
where $U\subseteq N$ is the set of unary nonterminals. Note that since $G$ is
partitioned, all unary symbols $A$ have $\iota(A)=(i,i)$ with $a_i\notin D$.

\begin{lemma}\label{plang}
For each partitioned grammar $G$, one can construct a breadth-bounded grammar
$G'$ with $\Lang{G'}=\PLang{G}$.
\end{lemma}
\begin{proof}
Suppose $G=(N,T,I,P,S)$ is a partitioned grammar with direct symbols
$D\subseteq T$ and with $\Lang{G}\subseteq a_1^*\cdots a_n^*$. Let $U\subseteq
N$ be the set of unary nonterminals.  We will use the new terminal symbols
$\bar{I}=\{\bar{f} \mid f\in I\}$ and $\bar{U}=\{\bar{A} \mid A\in U\}$.  If
$h\colon (U\cup I \cup T)^*\to (\bar{U}\cup \bar{I}\cup T)^*$ is the morphism
such that $h(a)=a$ for $a\in T$ and $h(x)=\bar{x}$ for $x\in U\cup I$, then it
clearly suffices to construct a breadth-bounded grammar $G'$ with
$\Lang{G'}=h(\PLang{G})$. Hence, our grammar will be of the form $G'=(N',
\bar{U}\cup\bar{I}\cup T, I, P', S)$.

The new set of nonterminals is $N'=N\cup \{Z\}$ for some fresh symbol $Z$.  The
productions of $G'$ are obtained as follows. First, we remove from $G$ all
productions where the nonterminal on the left-hand side is in $U$. Then, we add
for each $A\in U$ the production $A\to \bar{A}Z$ and for each $f\in I$ the
production $Zf\to \bar{f}Z$ and $Z\to\emptyWord$. Hence, the new productions
just output the nonterminal and then the index word (over a disjoint alphabet). 

It remains to be shown that $G'$ is breadth-bounded. Let $u$ be a sentential
form of $G'$.  Since $G$ is partitioned, we have $|u|_{U\cup \{Z\}}\le n$.
Moreover, there is a sentential form $v$ of $G$ with $|v|_{N\setminus
U}=|u|_{N\setminus U}$.  Suppose $A_1,\ldots,A_m$ are the non-unary
nonterminals in $v$. Since $G$ is an interval grammar, we have
$\iota(A_i)=(r_i, s_i)$ for $1\le i\le m$ such that $1\le r_1$, $s_m\le n$,
$r_i<s_i$ for $1\le i\le m$, and $s_i\le r_{i+1}$ for $1\le i<m$. This implies
$m\le n$. Therefore, $|u|_{N\setminus U}\le|v|_{N\setminus U}\le n$. Hence, we
have $|u|_{N'}\le 2n$. This proves that $G'$ is breadth-bounded.
\end{proof}

We are now ready to prove \Cref{partitionedbreadthbounded}.
\begin{proof}[Proof of \Cref{partitionedbreadthbounded}]
Suppose $G=(N,T,I,P,S)$ and $D\subseteq T$ is the set of direct symbols.
Without loss of generality, we assume that $T=\{a_1,\ldots,a_n\}$ and
that for some $m\le n$, we have $D=\{a_1,\ldots,a_m\}$. First, we
use \Cref{plang} to construct a breadth-bounded grammar $G''$ with
$\Lang{G''}=\PLang{G}$. This means $\Lang{G''}$ consists of words
\[ a_1^{x_1}\cdots a_m^{x_m}B_1w_1\cdots B_{n-m}w_{n-m},  \]
where $B_i\in N$, $w_i\in I^*$, and $\iota(B_i)=(m+i,m+i)$ for
$1\le i\le n-m$.

Let $A$ be the transducer provided by \Cref{transduction} with $f_A\approx
f_{G}$. We may clearly assume that $A$ always outputs words in $a^*$ for some
$a\in T$.  From $A$, we construct the transducer $A'$ that, on the input word
\[ a_1^{x_1}\cdots a_m^{x_m}B_1w_1\cdots B_{n-m}w_{n-m}, \]
$B_1,\ldots,B_{n-m}\in N$, $w_1,\ldots,w_{n-m}\in I^*$, outputs all those words 
\[ a_1^{x_1}\cdots a_m^{x_m}a_{m+1}^{y_1}\cdots a_n^{y_{n-m}} \]
for which $(B_{i}w_{i}, a^{y_i})\in \Trans{A}$ for $1\le i\le n-m$.

Let $V=\Trans{A'}$. According to \Cref{breadthboundedfulltrio}, we can compute
a breadth-bounded grammar $G'$ for $V(\PLang{G})=V(\Lang{G''})$. We claim that
$\Dclosure{\Lang{G'}}=a_1^*\cdots a_n^*$ if and only if
$\Dclosure{\Lang{G}}=a_1^*\cdots a_n^*$. 

Suppose $\Dclosure{\Lang{G}}=a_1^*\cdots a_n^*$. Then for each $k\in\N$, there
is a word $a_1^{x_{1,k}}\cdots a_n^{x_{n,k}}$ in $\Lang{G}$ such that
$x_{i,k}\ge k$.  This means, there are words 
\[ a_1^{x_{1,k}}\cdots a_m^{x_{m,k}}B_{1,k}w_{1,k}\cdots B_{n-m,k}w_{n-m,k}\in\PLang{G} \]
such that $x_{i,k}\ge k$ for $1\le i\le m$ and $f_G(B_{i,k}w_{i,k})\ge k$ for $1\le i\le n-m$.
Since $f_A\approx f_G$, this sequence of words has a subsequence
\[ a_1^{x'_{1,k}}\cdots a_m^{x'_{m,k}}B'_{1,k}w'_{1,k}\cdots B'_{n-m,k}w'_{n-m,k} \]
such that $f_A(B'_{1,k}w'_{1,k})\ge k$ for each $k\in\N$. Since this is a
subsequence, we still have $x'_{i,k}\ge k$ for $1\le i\le m$ and
$f_G(B'_{i,k}w'_{i,k})\ge k$ for $2\le i\le n-m$.  If we repeat this picking of
subsequences another $n-m-1$ times, we arrive at a sequence of words
\[ a_1^{\bar{x}_{1,k}}\cdots a_m^{\bar{x}_{m,k}}\bar{B}_{1,k}\bar{w}_{1,k}\cdots \bar{B}_{n-m,k}\bar{w}_{n-m,k} \]
in which $\bar{x}_{i,k}\ge k$ for $1\le i\le m$ and
$f_A(\bar{B}_{i,k}\bar{w}_{n-m,k})\ge k$ for $1\le i\le n-m$.  By definition of
$G'$, this yields words 
\[ a_1^{\bar{x}_{1,k}}\cdots a_m^{\bar{x}_{m,k}}a_{m+1}^{y_{1,k}}\cdots a_n^{y_{n-m,k}}\in\Lang{G'} \]
such that $y_{i,k}\ge k$ for $1\le i\le n-m$. Hence,
$\Dclosure{\Lang{G'}}=a_1^*\cdots a_n^*$. It can be shown completely
analogously that $\Dclosure{\Lang{G'}}=a_1^*\cdots a_n^*$ implies
$\Dclosure{\Lang{G}}=a_1^*\cdots a_n^*$.
\end{proof}

\subsection{Semilinearity}\label{step:semilinearity}
We have thus reduced the SUP for indexed grammars to the special case of
breadth-bounded grammars.  The last step in our proof is to prove the
following.  It clearly implies decidability of the SUP.
\begin{proposition}\label{breadthbounded}
Languages generated by breadth-bounded grammars have effectively
semilinear Parikh images.
\end{proposition}
The basic idea of \Cref{breadthbounded} is to use a decomposition of derivation
trees into a bounded number of `slices', which are edge sequences of either
\begin{enumerate*}[label=(\roman*)]\item only push and output productions
(`positive slice') or \item only pop and output productions (`negative
slice')\end{enumerate*}. Furthermore, there is a relation between slices such
that the index symbols that are pushed in a positive slice are popped precisely
in those negative slices related to it. One can then mimic the grammar by
simulating each positive slice in lockstep with all its related negative
slices.  This leads to a `finite index scattered context grammar'. This type of
grammars is well known to guarantee effectively semilinear Parikh
images~\cite{DassowPaun1989}.

\newcommand{\irreducible}{two-phased}
\newcommand{\irredGrammar}{two-phased}
\newcommand{\anIrreducible}{a}
\newcommand{\anIrredGrammar}{a}

Suppose $G$ is a breadth-bounded indexed grammar. Since $G$ can
be brought into normal form while preserving the property of
breadth-boundedness, we assume $G$ to be in normal form. Let $t$ be a
derivation tree for $G$. An edge in $t$ that connects nodes $x$ and $y$
is called \emph{chain edge} if $x$ and $y$ both have a label in $NI^*$
and $y$ is the only child of $x$ that has a label in $NI^*$. A non-empty
sequence of chain edges that forms a path is called a \emph{chain}. A
maximal chain (i.e. that cannot be prolonged on either side) is called a
\emph{segment}. An edge in $t$ corresponding to a push, pop, or output
production is called a \emph{push edge}, \emph{pop edge}, or output
edge, respectively. A chain that contains only push and output edges
is called a \emph{(positive) phase}. Similarly, if a chain contains
only pop and output edges, it is a \emph{(negative) phase}. For an
example of a derivation tree and the decomposition into segments and
phases, see~\Cref{derivationtreeslices}. The figure also shows an arrow
collection and the decomposition of phases into slices; these concepts
will be defined later.

We call a segment \emph{two-phased} if it consists of a negative
phase followed by a positive phase. In other words, this requires
that in the segment, there is no pop production applied anywhere
below a push production. For example, the derivation tree in
\Cref{derivationtreeslices} has only two-phased segments. If every
segment in every derivation tree of a grammar $G$ is two-phased, we say
that $G$ is \emph{two-phased}. Moreover, we call an indexed grammar
\emph{quasi-left-linear} if every output production is of the form $A\to
vB$, i.e. terminal words are only output on the left.

\begin{lemma}\label{irreducible}
For each breadth-bounded grammar $G$, one can construct a Parikh-equivalent
breadth-bounded grammar $G'$ that is \irredGrammar{}  and quasi-left-linear.
\end{lemma}
\begin{proof}
Let $G=(N,T,I,P,S)$ be breadth-bounded. First of all, by replacing each
production $A\to uBv$, $A,B\in N$, $u,v\in T^*$, with the production $A\to
uvB$, we obtain a Parikh-equivalent quasi-left-linear grammar. Hence, we may
assume that $G$ is quasi-left-linear.  We will use the restricted derivation
relation $\grammarsteplin[G]$, which requires that the applied productions are
part of a segment. This means, we have $x\grammarsteplin[G]y$ if $y$ is
obtained from $x$ by applying a push, pop, or output production.

We exploit the fact that derivations as above are essentially runs in a
pushdown automaton: The nonterminal can be regarded as a state, its index as a
stack content, and the generated terminal words correspond to the input of the
automaton. In particular, for each $A,B\in N$, the language
\[ L_{A,B}=\{ u\in T^* \mid A\grammarstepslin[G] uB \} \]
is context-free. Hence, we can construct a finite automaton $C_{A,B}$ whose language is
Parikh-equivalent to $L_{A,B}$.
Using the automata $C_{A,B}$, we will construct the new grammar $G'$.

The grammar $G'$ is obtained as follows. We assume that the state sets
of all the automata $C_{A,B}$ are pairwise disjoint and add each of
their states as a new nonterminal. For each edge $\autedge{p}{a}{q}$,
we add the production $p\to aq$. Moreover, we add the production $A\to
q_0$ for the initial state $q_0$ of $C_{A,B}$ and a production $f\to
B$ for each final state $f$ of $C_{A,B}$. Of course, since $G$ is
breadth-bounded, $G'$ is as well.

We clearly have $\Parikh{\Lang{G'}}=\Parikh{\Lang{G}}$ and we shall
prove that for each $w\in\Lang{G'}$, there is a $w'\in\Lang{G'}$
that satisfies $\Parikh{w'}=\Parikh{w}$ and can be derived using
only two-phased segments. The latter property will be refered to as
\emph{two-phase completeness}. We call two segments \emph{equivalent}
if they have the same initial and final nonterminal, the same effect
on the index, and generate terminal words with the same Parikh image.
Suppose $w\in\Lang{G'}$. We choose for $w$ a derivation tree $t$ for
$G'$ such that in each segment, the number of push or pop productions is
minimal among all equivalent segments. In other words, no segment in $t$
can be replaced by an equivalent one so that the number of push or pop
productions in this segment strictly decreases. We claim that then $t$
has only two-phased segments.

Suppose $t$ had a segment that is not two-phased. This means, it
contains a push production and, somewhere below, a pop production.
Then, somewhere in between, there is a push production, followed by
some output productions and a matching pop production. Here, `matching'
means that the pop production removes the index symbol added by the push
production. Let $A$ be the nonterminal to which the push production
is applied and let $B$ be the nonterminal that results from the pop
production. Since push and pop productions involve only nonterminals
that are already present in $G$, this means $A,B\in N$. We can therefore
replace the chain between the $A$-node and the $B$-node by productions
simulating a run of $C_{A,B}$. This strictly reduces the number of push
or pop productions and thus contradicts the choice of $t$. Thus, every
$w\in\Lang{G'}$ can be derived using derivation trees where all segments
are two-phased.

We can now easily turn $G'$ into a breadth-bounded grammar $G''$ that
does not allow segments that are not two-phased. This can be achieved
by endowing the nonterminals of $G'$ with an extra bit that indicates
whether the current segment already contains a push production. If
the latter is the case, no pop production is allowed for the rest of
the segment. Then, because of the two-phase completeness, $G''$ is
equivalent to $G'$ and hence Parikh-equivalent to $G$. Clearly, $G''$
inherits breadth-boundedness from $G'$ and is two-phased.
\end{proof}

We can now prove \Cref{breadthbounded} by showing that every 
quasi-left-linear breadth-bounded grammar can be turned into an equivalent 
grammar from a class for which effective semilinearity is well-known.
This type of grammar is called `finite index scattered context grammar'.

A \emph{scattered context grammar} is a tuple $G=(N,T,P,S)$, in which $N$ and
$T$ are disjoint alphabets of \emph{nonterminal} and \emph{terminal symbols},
respectively, $S\in N$ is the start symbol, and $P$ is a finite set of
sequences
\[ (A_1\to w_1, \ldots, A_n\to w_n) \]
of context-free productions, i.e. $A_i\in N$
and $w_i\in (N\cup T)^*$ for $1\le i\le n$. We apply a sequence by applying
all its productions in parallel to the current sentential form. Formally, we
have  $x\grammarstep[G] y$ if there is a production sequence $(A_1\to w_1,\ldots,A_n\to
w_n)\in P$ and a permutation $\pi$ of $\{1,\ldots,n\}$ such that
\begin{align*}
x=x_0A_{\pi(1)}x_1\cdots A_{\pi(n)}x_n, && y=x_0 w_{\pi(1)} x_1\cdots w_{\pi(n)} x_n
\end{align*}
for some $x_0,\ldots,x_n\in (N\cup T)^*$. The language \emph{generated by $G$} is then
\[ \Lang{G}=\{w\in T^* \mid S\grammarsteps[G] w \}.\]
The grammar $G$ is said to
have \emph{finite index} if there is a number $B\in\N$ such that each $w\in\Lang{G}$
has a derivation $S\grammarstep[G] w_1 \grammarstep[G]\cdots\grammarstep[G]
w_n=w$ such that $|w_i|_N\le B$ for each $1\le i\le n$. It is well-known (and not
hard to see) that languages generated by finite index scattered context
grammars are effectively semilinear~\cite{DassowPaun1989}.

\begin{lemma}\label{fisc}
Given a \irredGrammar{} quasi-left-linear breadth-bounded grammar, one can
construct an equivalent finite index scattered context grammar.
\end{lemma}

Our proof of \Cref{fisc} requires the decomposition of derivation trees into slices.

\newcommand{\edgelabel}[3]{#1.#2.#3}
\begin{figure}
	\begin{tikzpicture}[
	   sibling distance=15pt, 
	   level 1/.style={level distance=40pt},
	   level 2/.style={level distance=40pt},
	   level 3/.style={level distance=40pt},
	   level 5/.style={level distance=40pt},
	   level 6/.style={level distance=40pt},
	   level 4/.style={sibling distance=50pt}, 
	   level 6+/.style={sibling distance=0pt}
	   ]

		\clip (-5.5cm, -13.5cm) rectangle (7cm,1cm);
                \Tree [ .{$S$}
		        \edge node[midway] (1 i a 1) {} node[very near start, auto=left]{\edgelabel{1}{i}{$\alpha$}};
                        [ .{$Sf$}
		          \edge node[midway] (1 i a 2) {} node[very near start, auto=left]{\edgelabel{1}{i}{$\alpha$}};
                          [ .{$Sgf$}
		            \edge node[midway] (1 i b) {} node[very near start, auto=left]{\edgelabel{1}{i}{$\beta$}};
                            [ .{$Shgf$}
			      \edge[dashed];
                              [ .{$Ahgf$}
		                \edge node[midway] (2 ii c) {} node[auto=left]{\edgelabel{2}{ii}{$\gamma$}};
			        [ .{$Agf$}
		                  \edge node[midway] (2 iii d) {} node[very near start, auto=left]{\edgelabel{2}{iii}{$\delta$}};
                                  [ .{$Ahgf$}
				    \edge[dashed];
                                    [ .{$Ahgf$}
		                      \edge node[midway] (3 iv e) {} node[auto=left]{\edgelabel{3}{iv}{$\epsilon$}};
                                      [ .{$Agf$}
		                        \edge node[midway] (3 iv f 1) {} node[auto=left]{\edgelabel{3}{iv}{$\zeta$}};
                                        [ .{$Af$}
		                          \edge node[midway] (3 iv f 2) {} node[auto=left]{\edgelabel{3}{iv}{$\zeta$}};
					  [ .{$A$}
					    \edge[dashed];
                                            [ .{$a$}
                                            ]
					  ]
                                        ]
                                      ]
                                    ]
				    \edge[dashed];
                                    [ .{$Bhgf$}
		                      \edge node[midway] (4 v g) {} node[auto=right]{\edgelabel{4}{v}{$\eta$}};
                                      [ .{$Bgf$}
		                        \edge node[midway] (4 v h 1) {} node[auto=right]{\edgelabel{4}{v}{$\theta$}};
                                        [ .{$Bf$}
		                          \edge node[midway] (4 v h 2) {} node[auto=right]{\edgelabel{4}{v}{$\theta$}};
					  [ .{$B$}
					    \edge[dashed];
                                            [ .{$b$}
                                            ]
					  ]
                                        ]
                                      ]
                                    ]
                                  ]
				]
                              ]
			      \edge[dashed];
                              [ .{$Bhgf$}
			        \edge[dashed];
                                [ .{$c$}
                                ]
                                \edge node[midway] (5 vi i) {} node[auto=right]{\edgelabel{5}{vi}{$\iota$}};
                                [ .{$Chgf$}
                                  \edge[dashed];
                                  [ .{$Chgf$}
                                    \edge node[midway] (6 vii j) {} node[auto=left]{\edgelabel{6}{vii}{$\kappa$}};
                                    [ .{$Cgf$}
                                      \edge node[midway] (6 vii k 1) {} node[auto=left]{\edgelabel{6}{vii}{$\lambda$}};
                                      [ .{$Cf$}
                                        \edge node[midway] (6 vii k 2) {} node[auto=left]{\edgelabel{6}{vii}{$\lambda$}};
                                        [ .{$C$}
                                          \edge[dashed];
                                          [ .{$c$}
                                          ]
                                        ]
                                      ]
                                    ]
                                  ]
				  \edge[dashed];
                                  [ .{$Dhgf$}
				    \edge node[midway] (7 viii l) {} node[auto=right]{\edgelabel{7}{viii}{$\mu$}};
                                    [ .{$Dgf$}
				      \edge node[midway] (7 viii m 1) {} node[auto=right]{\edgelabel{7}{viii}{$\nu$}};
                                      [ .{$Df$}
				        \edge node[midway] (7 viii m 2) {} node[auto=right]{\edgelabel{7}{viii}{$\nu$}};
                                        [ .{$D$}
					  \edge[dashed];
                                          [ .{$d$}
                                          ]
                                        ]
                                      ]
                                    ]
                                  ]
                                ]
                              ]
                            ]
                          ]
                        ]
                      ]
	\draw[->] (1 i a 1) .. controls +(180:3.5cm) and +(180:3.5cm) .. (3 iv f 2);
	\draw[->] (1 i a 2) .. controls +(180:3cm) and +(180:3cm) .. (3 iv f 1);
	\draw[->] (1 i a 1) .. controls +(180:1.7cm) and +(0:1.7cm) .. (4 v h 2);
	\draw[->] (1 i a 2) .. controls +(180:1cm) and +(0:2cm) .. (4 v h 1);
	\draw[->] (1 i a 1) .. controls +(0:2.2cm) and +(180:1.5cm) .. (6 vii k 2);
	\draw[->] (1 i a 2) .. controls +(0:2.2cm) and +(180:2cm) .. (6 vii k 1);
	\draw[->] (1 i a 1) .. controls +(0:3.5cm) and +(0:3.5cm) .. (7 viii m 2);
	\draw[->] (1 i a 2) .. controls +(0:3cm) and +(0:3cm) .. (7 viii m 1);

	\draw[->] (1 i b) .. controls +(180:2cm) and +(180:2cm) .. (2 ii c);
	\draw[->] (1 i b) .. controls +(0:2cm) and +(0:2.1cm) .. (7 viii l);
	\draw[->] (1 i b) .. controls +(0:1cm) and +(180:2.5cm) .. (6 vii j);
	\draw[->] (2 iii d) .. controls +(180:1.4cm) and +(180:1.4cm) .. (3 iv e);
	\draw[->] (2 iii d) .. controls +(0:1.4cm) and +(0:1.4cm) .. (4 v g);
	\end{tikzpicture}
\caption{Derivation tree with arrow collection. Solid edges represent
chain edges. Each chain edge has a label \edgelabel{$x$}{$y$}{$z$},
where $x$, $y$, $z$ indicate the segment, phase, and slice to which the
edge belongs. 
$S,A,B,C$ are
nonterminal symbols, $f,g,h$ are index symbols, and $a,b,c,d$ are terminal
symbols.}\label{derivationtreeslices}
\end{figure}
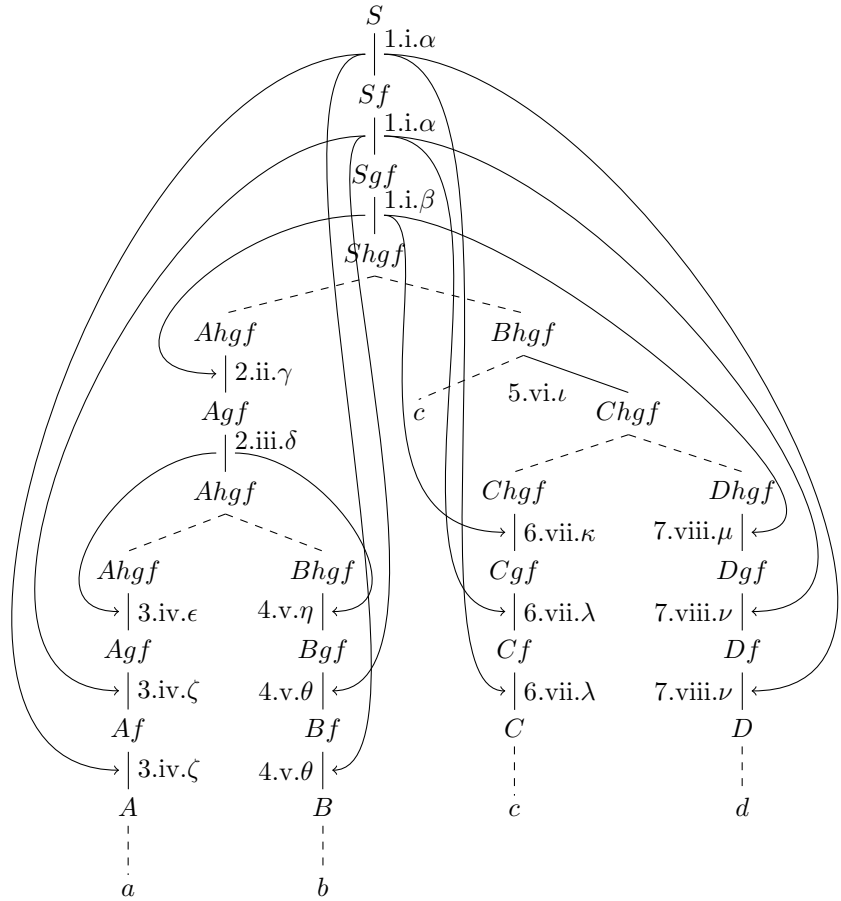

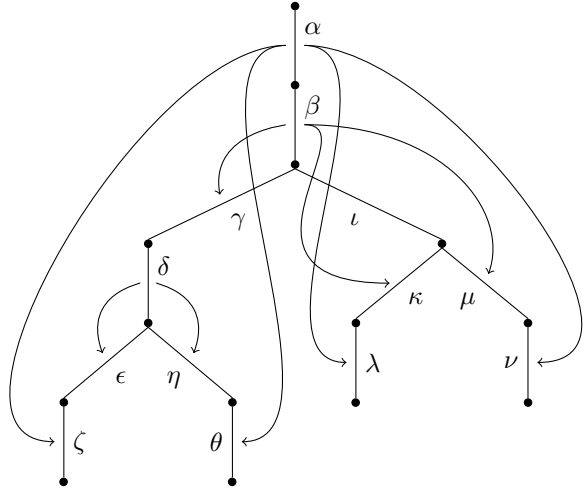
\begin{figure}
        \begin{tikzpicture}[
                sibling distance=40pt,
                level 3/.style={sibling distance=100pt},
                   slicenode/.style={circle,draw,fill=black,inner sep=1pt}
                ]
                \Tree [ .\node[slicenode] {};
                        \edge node[midway] (1 i a) {} node[near start, auto=left]{$\alpha$};
                        [ .\node[slicenode] {}; 
                          \edge node[midway] (1 i b) {} node[near start, auto=left]{$\beta$};
                          [ .\node[slicenode] {};
                            \edge node[midway] (2 ii c) {} node[auto=left]{$\gamma$};
                            [ .\node[slicenode] {};
                              \edge node[midway] (2 iii d) {} node[near start, auto=left]{$\delta$};
                              [ .\node[slicenode] {};
                                \edge node[midway] (3 iv e) {} node[auto=left]{$\epsilon$};
                                [ .\node[slicenode] {};
                                  \edge node[midway] (3 iv f) {} node[auto=left]{$\zeta$};
                                  [ .\node[slicenode] {};
                                  ]
                                ]
                                \edge node[midway] (4 v g) {} node[auto=right]{$\eta$};
                                [ .\node[slicenode] {};
                                  \edge node[midway] (4 v h) {} node[auto=right]{$\theta$};
                                  [ .\node[slicenode] {};
                                  ]
                                ]
                              ]
                            ]
                            \edge node[midway] (5 vi i) {} node[auto=right]{$\iota$};
                            [ .\node[slicenode] {};
			      \edge node[midway] (6 vii j) {} node[auto=left]{$\kappa$};
                              [ .\node[slicenode] {};
				\edge node[midway] (6 vii k) {} node[auto=left]{$\lambda$};
                                [ .\node[slicenode] {};
                                ]
                              ]
		              \edge node[midway] (7 viii l) {} node[auto=right]{$\mu$};
                              [ .\node[slicenode] {};
				\edge node[midway] (7 viii m) {} node[auto=right]{$\nu$};
                                [ .\node[slicenode] {};
                                ]
                              ]
                            ]
                          ]
                        ]
                      ]
	\draw[->] (1 i a) .. controls +(180:2cm) and +(180:2cm) .. (3 iv f);
	\draw[->] (1 i a) .. controls +(180:1.7cm) and +(0:1.7cm) .. (4 v h);
	\draw[->] (1 i a) .. controls +(0:1.5cm) and +(180:1.5cm) .. (6 vii k);
	\draw[->] (1 i a) .. controls +(0:2cm) and +(0:2cm) .. (7 viii m);

	\draw[->] (1 i b) .. controls +(180:1pt) and +(100:1cm) .. (2 ii c);
	\draw[->] (1 i b) .. controls +(0:2cm) and +(70:1cm) .. (7 viii l);
	\draw[->] (1 i b) .. controls +(0:1cm) and +(180:2.5cm) .. (6 vii j);
	\draw[->] (2 iii d) .. controls +(180:1pt) and +(110:1cm) .. (3 iv e);
	\draw[->] (2 iii d) .. controls +(0:1pt) and +(70:1cm) .. (4 v g);
        \end{tikzpicture}
\caption{Slice tree arising from the derivation tree in
\Cref{derivationtreeslices}. Each edge label indicates the slice that
induces the edge.}\label{slicetree}
\end{figure}

\subsection*{Slices}
Suppose $t$ is a tree with edge set $E$. An \emph{arrow collection for
$t$} is a finite set $A$ together with two maps $\nu_0,\nu_1\colon A\to
E$. If $\nu_0(a)=e$ and $\nu_1(f)$ for edges $e,f$ of $t$, we call $e$
and $f$ the \emph{source} and \emph{target} of $a$ and $a$ is an arrow
\emph{from $e$ to $f$}. If $t$ is a derivation tree of a breadth-bounded
grammar $G$ such that all segments of $t$ are two-phased, we endow it
with an arrow collection: From each push edge $e$, we draw an arrow to
each of the pop edges that remove the index symbol created by $e$. If
$e$ is a push edge, its \emph{type} is the set of phases at which arrows
from $e$ arrive.

Since $G$ is breadth-bounded, we have an upper bound on the number
of segments in derivation trees: If $k$ is a bound on the number of
nonterminals in sentential forms, then a derivation can contain at
most $k-1$ split productions; and if there are at most $k-1$ split
productions, a derivation tree can contain at most $2(k-1)+1$ segments.
Since $G$ is two-phased, this entails a bound on the number of phases
in derivation trees. In particular, the number of types of push edges
is bounded as well. A positive phase in which all push edges have equal
type is called a \emph{(positive) slice}. Observe that if in some
positive phase, the push edges $e$ and $f$ have the same type, then
every push edge between $e$ and $f$ must also have this type. This means
each positive phase decomposes into a bounded number of positive slices.

Consider a derivation tree $t$ for $G$ and a decomposition of each
segment into $\le 2$ phases. In the same way, we assume a decomposition
of each positive phase in $t$ into positive slices. Note that each pop
edge is connected by an arrow to a unique push edge. Therefore, the
\emph{type} of a pop edge is the positive slice containing this push
edge. A negative phase in which all pop edges have equal type is called
a \emph{(negative) slice}. As above, we can argue that each negative
phase decomposes into a bounded number of negative slices. For an example
of a derivation tree and its decomposition into segments, phases, and slices,
see~\Cref{derivationtreeslices}.

By a simple modification to $G$, we may assume that there is a chain edge
\begin{enumerate*}[label=(\roman*)]\item directly below the root node and \item
directly below each node created by a split production.\end{enumerate*} In
other words, at the beginning of the derivation as well as directly below each
node created by a split production, a segment begins.

\subsection*{Slice trees} The decomposition of segments into a bounded number of
slices gives rise to the concept of slice trees. A \emph{slice tree} is a tree
together with an arrow collection $A$, such that the following holds:
\begin{enumerate}[label=(\roman*)]
\item For each arrow $a$, $\nu_0(a)$ is an ancestor of $\nu_1(a)$ (in other
words, there is a path from the root to a leaf that contains $\nu_0(a)$ and
$\nu_1(a)$ such that $\nu_0(a)$ is closer to the root).
\item 
   Every edge $e$ is either \begin{enumerate*}[label=(\arabic*)] \item
   a \emph{positive edge}, meaning that there is at least one arrow
   leaving $e$ and no arrow arriving in $e$ or \item a \emph{negative
   edge}, meaning that there is precisely one arrow arriving in $e$
   and no arrow leaving in $e$, or \item there is no arrow leaving or
   arriving in $e$. \end{enumerate*}
\item If $e$ is a positive edge, then for every path from $e$ to a leaf, there is
an arrow from $e$ arriving on this path.
\item On each path from the root to a leaf, the arrows are \emph{well-nested},
meaning there is no subsequence of edges $e, f, g, h$ and an arrow from $e$ to
$g$ and an arrow from $f$ to $h$.
\end{enumerate}

To each derivation tree $t$ of $G$, we associate a slice trees $\bar{t}$
as follows. We choose a decomposition of $t$'s segments into phases
and then a decomposition of phases into slices. We delete all nodes
with label in $T\cup\{\emptyWord\}$ (in other words, all leaves) and
we merge each slice (positive or negative) down to a single edge.
Now, the only edges that do not result from merging a slice are those
created by split productions. Because of our modification, there is a
slice edge directly below, so that we can merge them with this slice
edge below. The arrows in $\bar{t}$ arise from the arrows in $t$: If
there are arrows from one slice to another in $t$, then we add an
arrow between the corresponding edges in $\bar{t}$. This completes the
description of the slice tree $\bar{t}$. As an example, the derivation
tree in \Cref{derivationtreeslices} results in the slice tree in
\Cref{slicetree}.

Note that there is a one-to-one correspondence between the edges of the
slice tree $\bar{t}$ and the slices of $t$. Moreover, the branching
nodes of $\bar{t}$ correspond to applications of split productions
in $t$; and degree-one nodes in $\bar{t}$ correspond to nodes that
are incident to two slices. Since the edges in $\bar{t}$ are in
correspondence with the slices of $t$, we also call them \emph{slices}.
Furthermore, since we have seen above that we have a bound on the number
of slices in a derivation tree for $G$, we have an upper bound for the
size of all slice trees of derivation trees for $G$.

We are now ready to prove \Cref{fisc}.

\begin{proof}[Proof of \Cref{fisc}]
We may assume that every terminal production in $G$ is of the form
$A\to\emptyWord$.  Fix a slice tree $t$.  We will construct a finite index
scattered context grammar that generates all words in $\Lang{G}$ that have a
derivation tree with slice tree $t$. Since we have a bound on the size of slice
trees for derivation trees of $G$ and the languages generated by finite index
scattered context grammars are closed under finite unions, this clearly
suffices.

\newcommand{\Initial}[1]{X_{#1}}
\newcommand{\Target}[1]{Y_{#1}}
\newcommand{\PosNeg}[1]{Z_{#1}}
Consider a derivation with slice tree $t$. Then each slice $s$ starts in a
certain nonterminal $\Initial{s}$ and ends in a nonterminal $\Target{s}$. These
satisfy the following conditions:
\begin{enumerate}[label=(S\arabic*)]
\item\label{ntcond:connect} If a slice $s_2$ is a direct descendant of $s_1$,
then $\Initial{s_2}=\Target{s_1}$.
\item\label{ntcond:split} If $x$ is a node with two children, $s$ is the slice
above $x$ and $s_1,s_2$ are the slices below $x$, then there is a production
$\Target{s}\to \Initial{s_1}\Initial{s_2}$ in $G$.
\item\label{ntcond:start} If $s$ is the slice below $t$'s root (note that $t$'s
root has degree $1$), then $\Initial{s}$ is the start symbol $S$ of $G$.
\item\label{ntcond:terminal} If $s$ is a slice whose lower node is a leaf of
$t$, then there is a production $\Target{s}\to\emptyWord$ in $G$.
\end{enumerate}
Since there are only finitely many ways to choose the nonterminals
$\Initial{s}$ and $\Target{s}$ for each slice $s$, we may assume a fixed choice
such that \labelcref{ntcond:connect,ntcond:split,ntcond:terminal,ntcond:start} are
satisfied and construct a grammar that simulates all derivations in which this
choice occurs.

The nonterminals of our grammar $G'$ are pairs $(s,A)$, where $s$ is a slice
and $A$ is a nonterminal of $G$.  The idea behind the construction of $G'$ is
that each sentential form contains one pair $(s,A)$ for each slice $s$. This
nonterminal creates the same output as the slice $s$ in the derivation of $G$.
In order to simulate the index words, we have production sequences that
simulate each push production in parallel to all matching pop productions.
This is possible since all the index symbols pushed in some positive slice $s$
are popped in those negative slices $s_1,\ldots,s_n$ for which there are arrows
from $s$ to $s_1,\ldots,s_n$.

Hence, positive slices are simulated in the same order as their productions are
applied in $G$ and negative slices are simulated in reverse. Therefore, we
define $\PosNeg{s}=\Initial{s}$ if $s$ is a positive slice and
$\PosNeg{s}=\Target{s}$ if $s$ is a negative slice.

The grammar $G'$ begins each derivation by
producing a string $f_t$ that is defined inductively by describing $f_u$ for
every subtree $u$ of $t$. Let $r$ be the root node of $u$. If $r$ has no
children, then $f_u=\emptyWord$.  If $r$ has one child, then we denote the
subtree under $r$'s child node by $u'$ and define  $f_u=(s,
\PosNeg{s})f_{u'}$, where $s$ is the slice incident to $r$. If $r$ has two
children $c_1$ and $c_2$, then we denote the trees under $c_1$ and $c_2$ by
$u_1$ and $u_2$ and define 
\[ f_u=(s_1, \PosNeg{s_1})f_{u_1}(s_2,\PosNeg{s_2})f_{u_2}. \]
In other words, we perform a pre-order traversal and when we walk along a slice
$s$, we append $(s,\PosNeg{s})$ on the right.

Let us describe the production sequences in $G'$. First of all, we have a
sequence that produces the word $f_t$, namely $(S\to f_t)$. In order to
simulate the creation of a stack symbol in a positive slice $s$ that is
consumed in $s$'s corresponding negative slices $s_1,\ldots,s_n$, we have the
sequence
\[ ((s, A) \to (s, B), (s_1, B_1)\to (s_1, A_1), \ldots (s_n, B_n) \to (s_n, A_n)) \]
for each $f\in I$ such that there are productions $A\to Bf$ and $A_if\to B_i$
in $G$ for $A,B\in N$, and $A_i,B_i\in N$ for $1\le i\le n$.  In order to
simulate the output production $A\to vB$, we have a sequence $((s, A)\to
v(s,B))$ for each positive slice $s$ and a sequence $((s, B)\to (s,
A)v)$ for each negative slice $s$.

Finally, we need sequences that remove the nonterminals if they match the
target (initial) nonterminal chosen for the respective positive slice (negative
slice). Hence, we add the sequence $((s,\Target{s})\to\emptyWord)$ for each
positive slice $s$ and the sequence $((s,\Initial{s})\to\emptyWord)$ for each
negative slice $s$.  It is clear from the construction that then $\Lang{G'}=\Lang{G}$.
\end{proof}

\subsection*{Acknowledgements}
The author would like to thank Sylvain Schmitz, who pointed out to him
that \citeauthor{Jullien1969}~\cite{Jullien1969} was the
first to characterize downward closed languages by simple regular expressions.

\printbibliography

\end{document}